\documentclass[%
 reprint,
superscriptaddress,
 amsmath,amssymb,
 aps,
]{revtex4-2}

\usepackage{graphicx}% Include figure files
\usepackage{dcolumn}% Align table columns on decimal point
\usepackage{bm}% bold math
\usepackage{xcolor}
\usepackage{xr-hyper}
\usepackage{hyperref}% add hypertext capabilities
\hypersetup{colorlinks = true, linkcolor = blue!75!black, citecolor = blue!75!black, urlcolor = blue!75!black}
 
\begin{document}

\preprint{APS/123-QED}

\title{Core-excited states of SF$_{6}$ probed with soft X-ray femtosecond transient absorption of vibrational wavepackets}

\author{Lou Barreau}
\altaffiliation[Present address: ]{Institut des Sciences Moléculaires d’Orsay, UMR 8214, CNRS, Université Paris-Saclay, Bâtiment 520, 91405 Orsay Cedex, France.}
\email{lou.barreau@universite-paris-saclay.fr}
\affiliation{Department of Chemistry, University of California, Berkeley, CA, 94720, USA}
\affiliation{Chemical Sciences Division, Lawrence Berkeley National Laboratory, Berkeley, CA, 94720, USA}
\author{Andrew D. Ross}
\affiliation{Department of Chemistry, University of California, Berkeley, CA, 94720, USA}
\affiliation{Chemical Sciences Division, Lawrence Berkeley National Laboratory, Berkeley, CA, 94720, USA}
\author{{Victor Kimberg}}
\email{kimberg@kth.se}
\affiliation{Department of Theoretical Chemistry and Biology, KTH Royal Institute of Technology, 10691 Stockholm, Sweden}
\author{{Pavel Krasnov}}
\affiliation{International Research Center of Spectroscopy and Quantum Chemistry -- IRC SQC, Siberian Federal University, Krasnoyarsk, Russia 660041}
\author{{Svyatoslav Blinov}}
\affiliation{International Research Center of Spectroscopy and Quantum Chemistry -- IRC SQC, Siberian Federal University, Krasnoyarsk, Russia 660041}
\author{Daniel M. Neumark}
\email{dneumark@berkeley.edu}
\affiliation{Department of Chemistry, University of California, Berkeley, CA, 94720, USA}
\affiliation{Chemical Sciences Division, Lawrence Berkeley National Laboratory, Berkeley, CA, 94720, USA}
\author{Stephen R. Leone}
\email{srl@berkeley.edu}
\affiliation{Department of Chemistry, University of California, Berkeley, CA, 94720, USA}
\affiliation{Chemical Sciences Division, Lawrence Berkeley National Laboratory, Berkeley, CA, 94720, USA}
\affiliation{Department of Physics, University of California, Berkeley, CA, 94720, USA}

\date{\today}% It is always \today, today,
             %  but any date may be explicitly specified

\begin{abstract}
A vibrational wavepacket in SF$_6$, created by impulsive stimulated Raman scattering with a few-cycle infrared pulse, is mapped onto five sulfur core-excited states using table-top soft X-ray transient absorption spectroscopy between 170-200 eV. The amplitudes of the X-ray energy shifts of the femtosecond oscillations depend strongly on the nature of the state. The prepared wavepacket is controlled with the pump laser intensity to probe the core-excited levels for various extensions of the S-F stretching motion. This allows the determination of the relative core-level potential energy gradients, in good agreement with TDDFT calculations. This experiment demonstrates a new means of characterizing core-excited potential energy surfaces.
\end{abstract}

\maketitle

Molecular potential energy surfaces (PESs) dictate the coupled electron-nuclear dynamics following electronic excitation. In particular, PESs of core-excited states are of considerable interest because core-level excitation can induce ultrafast nuclear motion on a timescale shorter than the few-femtosecond core-hole lifetime \cite{GelmukhanovPRA1996,SimonPRL1997,MarchenkoPRL2017}. The X-ray absorption spectrum of a molecule in its vibrational ground state probes only a small region of the PES, at the equilibrium geometry of the ground electronic state. In order to access a larger range of the core-excited PES, multiple infrared (IR) pump - X-ray probe schemes have been theoretically proposed \cite{FelicissimoPRA2005,GuimaraesPRA2005,felicissimoJCP2005,carniatoCPL2007,enginCPL2012,IgnatovaPRA2017}; the IR pulse excites the molecule to higher vibrational states where the nuclear wavepacket has a larger spatial extension, so that subsequent absorption of the X-ray pulse can probe regions of the PES that are otherwise inaccessible. The experimental implementation of these proposals requires few-femtosecond to attosecond X-ray pulses, now available at X-ray Free Electron Laser facilities and from table-top sources based on high-order harmonic generation (HHG) \cite{YoungJPhysB2018}. Indeed, X-ray transient absorption spectroscopy is a sensitive probe of structural dynamics \cite{geneauxPhilTrans2019}; as the geometry changes, the energy of the electronic transition in the X-ray region is modified. This technique has been successfully used to observe vibrational wavepackets in neutral or cationic molecules, often accompanying strong-field ionization in e.g. Br$_2$ \cite{HoslerPRA2013,KobayashiPRA2020a}, DBr \cite{KobayashiPRA2020b}, NO \cite{saitoOptica2019}, CH$_3$I \cite{WeiNatComm2017}, CH$_3$Br \cite{timmersNatComm2019} and C$_2$H$_4$ \cite{zinchenkoScience2021}, or single-photon and Raman excitation in I$_2$ \cite{PoullainPRA2021} and alkyl iodides \cite{ChangJCP2022}. In these cases, the vibrational coherence is typically mapped onto a dissociative core-excited state of predominantly $nd^{-1}\sigma^{*}$ (for halogen-containing species) or $1s^{-1}\pi^{*}$ character, which corresponds in the single-particle picture to the excitation of a non-bonding core electron to an anti-bonding molecular orbital.

\begin{figure}%
    \centering
    \includegraphics[width=1\linewidth]{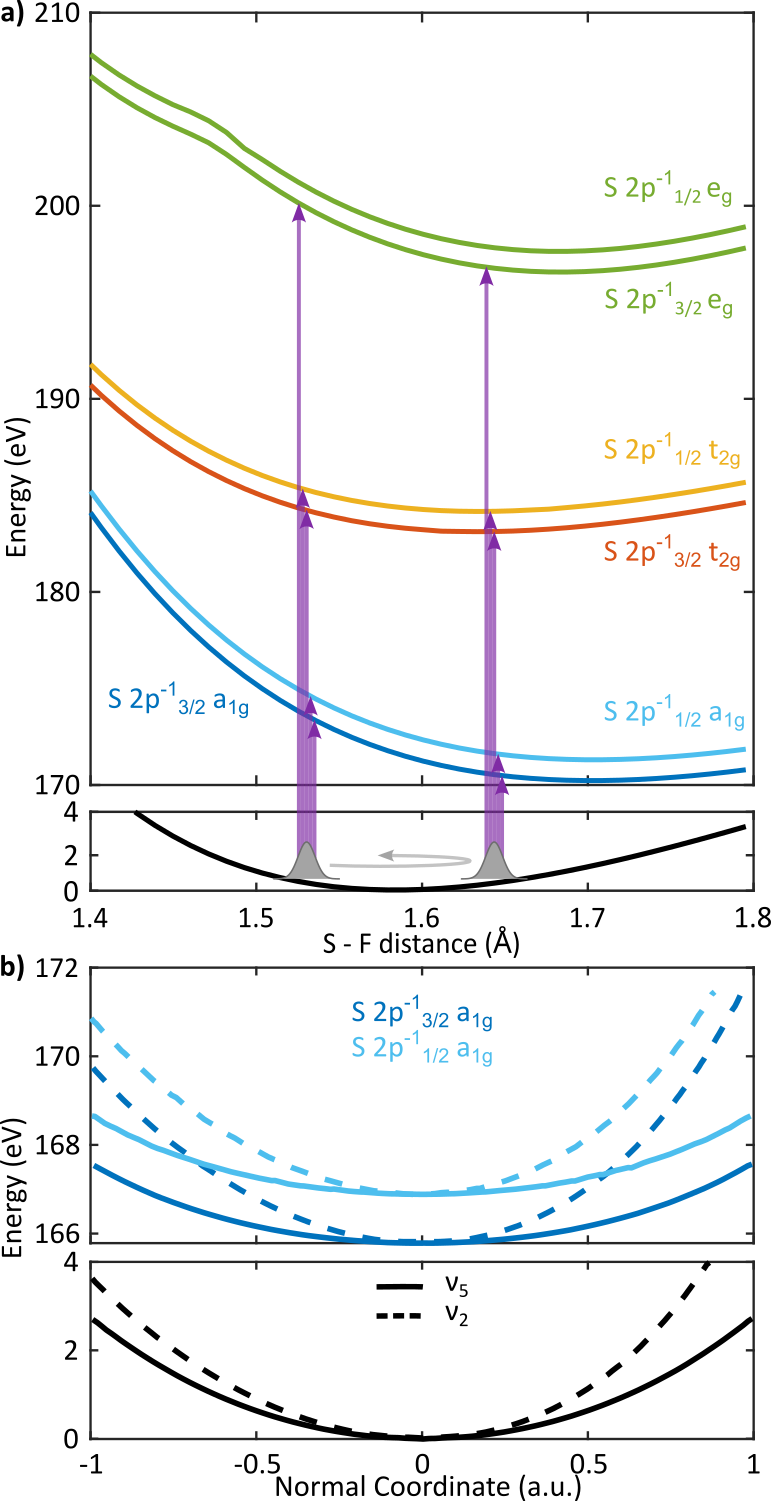}
    \caption{\textbf{a)} Illustrative schematic of the experiment principle overlaid on the calculated ground (bottom, black) and core-excited (top) states potential energy curves along the $\nu_1$ mode. \textbf{b)} Same as a) along the $\nu_2$ (dashed line) and $\nu_5$ (full line) modes. Only the 2p$^{-1}$~$a_{1g}$ states are shown because these vibrational modes lift the degeneracy of the 2p$^{-1}$~$t_{2g}$ and 2p$^{-1}$~$e_{g}$ states outside the equilibrium geometry.}
    \label{fig:Pot_v1}%
\end{figure}

\begin{figure}
    \centering
    \includegraphics[width=\linewidth]{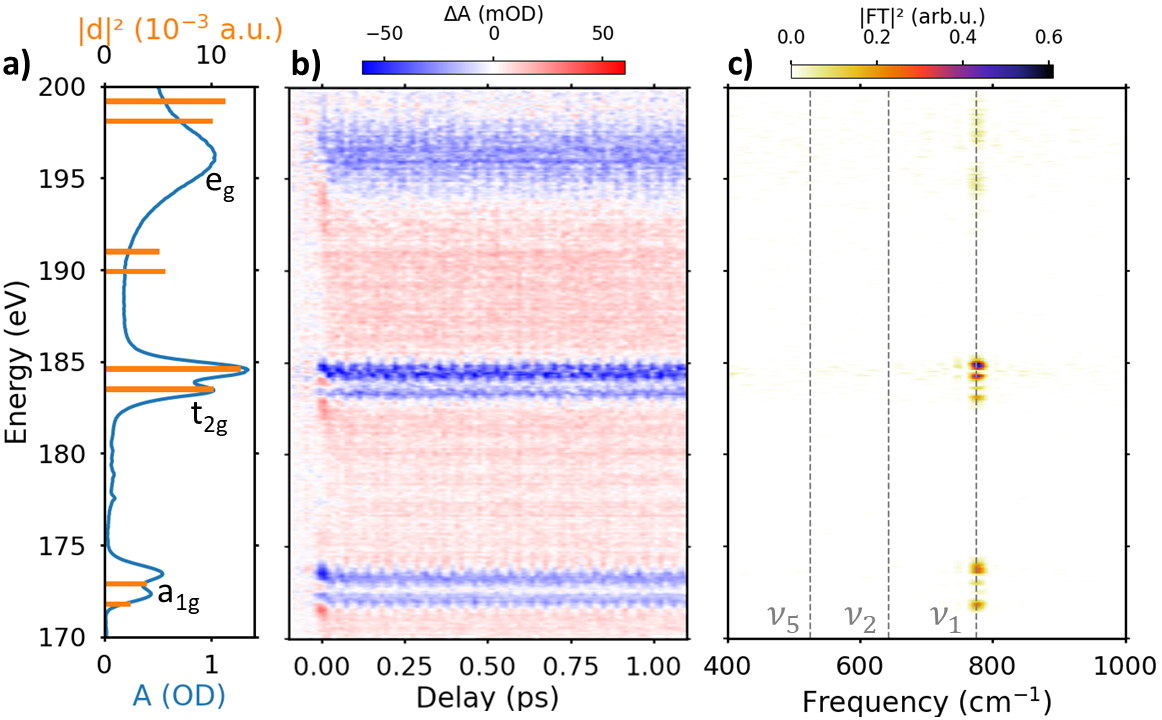}
   \caption{\textbf{a)} Measured (blue) and calculated (orange) SF$_6$ absorption spectrum at the S L$_{2,3}$ edge. The pre-edge background absorption has been subtracted in the measurement. The calculated energies are shifted by +6.0 eV. The additional doublet of $e_g$ symmetry calculated around 190 eV is not observed in the experiment. \textbf{b)} Transient absorption spectrogram. Positive delays correspond to the SXR pulse following the vis-NIR pulse. \textbf{c)} Squared Fourier transform of \textbf{b}. Frequencies of the $\nu_1$, $\nu_2$ and $\nu_5$ modes are indicated as vertical dashed lines.}
    \label{fig:Transient}%
\end{figure}

In this letter, we use a combination of  IR and soft X-ray (SXR) few-femtosecond pulses to experimentally map a vibrational wavepacket onto five sulfur L-core-excited states of SF$_6$ in the 170-200 eV energy range (Fig.~\ref{fig:Pot_v1}). The IR pump pulse produces a coherent superposition of vibrational states in the ground electronic state by Impulsive Stimulated Raman Scattering (ISRS) \cite{yanJCP1985,yanJCP1987} and the X-ray absorption energy is probed as a function of the time-delay between the two pulses. The amplitude of the oscillations in energy observed in the transient absorption depends strongly on the core-excited state, in agreement with the nature of the populated molecular orbitals and with Time-Dependent Density Functional Theory (TDDFT) calculations of the core-excited PESs. The intensity of the short IR pulse is used to control the number of vibrational states in the superposition to extract one-dimensional potential energy gradients along a normal mode selected by detection at its vibrational frequency \cite{gerberScience2017}. 

The SF$_6$ molecule is chosen for its numerous S 2p core-excited states accessible in the SXR, due to the presence of a shape resonance: 2p$^{-1}_{3/2}$~$a_{1g}$ and 2p$^{-1}_{1/2}$~$a_{1g}$ which lie below the respective 2p ionization potentials at 172.27 and 173.44 eV respectively, and 2p$^{-1}_{3/2}$~$t_{2g}$, 2p$^{-1}_{1/2}$~$t_{2g}$ and 2p$^{-1}$~$e_{g}$ located in the shape resonance at 183.40, 184.57 and 196.2 eV respectively \cite{HudsonPRA1993,stenerJCP2011}. This polyatomic molecule possesses fifteen vibrational modes but due to its high symmetry many of them are degenerate. Among those, three are Raman-active \cite{claassenJCP1970}: the symmetric stretch $\nu_1$ ($\nu_1$ = 775 cm$^{-1}$, T$_1$ = 43 fs), antisymmetric stretch $\nu_2$ ($\nu_2$ = 643 cm$^{-1}$, T$_2$ = 52 fs), and bend $\nu_5$ ($\nu_5$ = 525 cm$^{-1}$, T$_5$ = 63 fs). They can be excited by ISRS with IR pulses shorter than their vibrational period, and the corresponding multimode wavepacket dynamics have been characterized with high-harmonic spectroscopy \cite{wagnerPNAS2006,ferreJPB2014,ferreNatComm2015,BaykushevaPRL2016}. Figure~\ref{fig:Pot_v1} shows the relevant ground and core-excited potential energy curves calculated by, respectively, DFT \cite{HohenbergPR1964} and TDDFT \cite{RungePRL1984} methods using the ORCA software \cite{NeeseORCA} with B3LYP functional \cite{BeckeJCP1993,LeePRB1988,vosko1980}, ano-pVTZ basis set \cite{NeeseJCTC2011}, RI approximation \cite{KendallFruchtl1997} and quasi-degenerated perturbation theory \cite{deSouzaJCTC2019} for spin-orbit coupling inclusion. Preliminarily, the geometry and vibrational modes of SF$_6$ were calculated by DFT under O$_h$ symmetry constrained with the help of the GAMESS software \cite{Gamess} using the B3LYP functional and def2-TZVP basis set \cite{WeigendPCCP2005}.

The experimental setup has been described elsewhere \cite{BarreauSciRep2020}. Briefly, 85\% of the energy of a commercial Ti:sapphire laser delivering 13~mJ, 30~fs, 800~nm pulses is used to pump a multistage optical parametric amplifier that converts the wavelength to 1300 nm. The resulting short-wave infrared (SWIR) pulses are compressed to 12.8 fs full-width at half-maximum (fwhm) with a hollow-core fiber (HCF) compressor, and focused into a semi-infinite gas cell filled with 2 bar of flowing helium for HHG. This yields SXR pulses with a continuous spectrum extending up to 370~eV. The remaining SWIR light is filtered out with a Sn film, and the SXR pulses are focused by a toroidal mirror into a gas cell filled with 25~mbar of SF$_6$. The SXR spectrum $I$ is measured after dispersion on a grating and imaging on an X-ray CCD camera. The absorbance $A$ is defined as $A= -\log_{10}(I/I_0)$, where $I_0$ is the spectrum measured without sample. 

In the time-resolved experiments, the remaining 2~mJ, 800~nm pulses are simultaneously compressed in a second HCF compressor to produce 0.75~mJ, 6~fs pulses in the visible-near IR (vis-NIR). After propagation in a piezo-controlled delay line, the vis-NIR pulses are focused with a $f=$~37.5 cm mirror into the gas cell to excite the molecules. The change in absorbance at a delay $\tau$ after the pump vis-NIR pulse is $\Delta A (\tau) = -\log_{10}(I_\text{on} (\tau)/I_\text{off})$, where $I_\text{on}$ and $I_\text{off}$ are the spectra measured with and without the pump pulse at each delay, respectively. The ensemble of pump-off spectra is used for the edge-pixel referencing technique applied to reduce the SXR fluctuations noise in the transient absorption data \cite{GeneauxOE2021}. In this all-optical experiment, the time and energy resolutions are not interdependent through the uncertainty principle (as opposed to core-level photoelectron spectroscopy that was proposed in \cite{NguyenPRA2016} to probe SF$_6$ vibrations) so that they can be both optimally short and narrow, respectively.

\begin{figure}%
  \centering
    \includegraphics[width=0.85\linewidth]{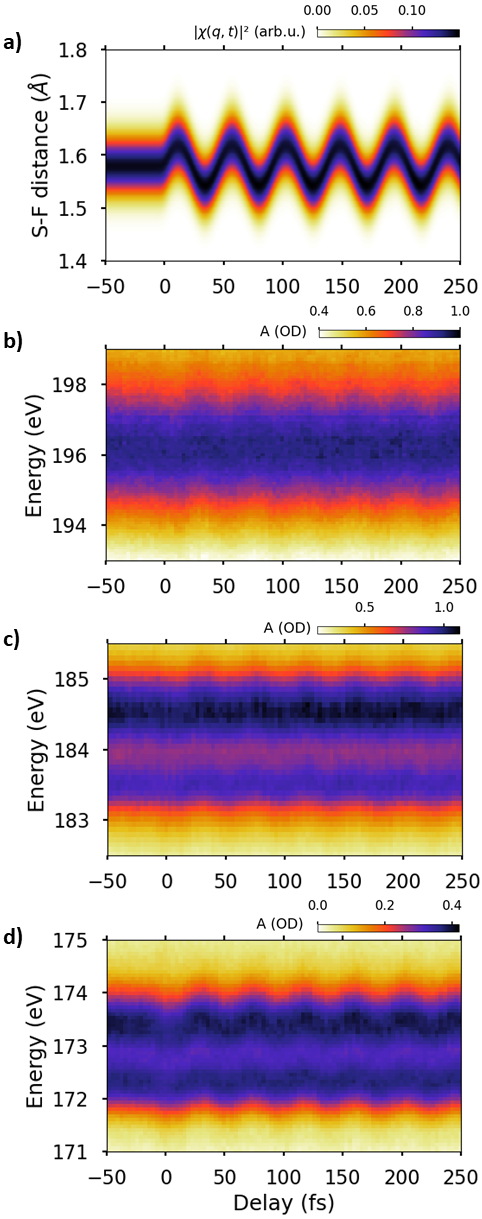}
   \caption{\textbf{a)} Simulated nuclear wavepacket after ISRS excitation of the $\nu_1$ mode by a 6~fs, 800~nm pulse of intensity $6 \times 10^{14}$ W/cm$^2$. \textbf{b-d)} Measured absorbance $A(\tau)$ for the five core-excited states at the same pump intensity.}
   \label{fig:Wavepacket}%
\end{figure}

Figure~\ref{fig:Transient}a shows the absorption spectrum of SF$_6$ in the vicinity of the S L-edge measured in the absence of the vis-NIR pulse (blue curve). Five peaks are observed, corresponding to the excitation to the spin-orbit split S 2p$^{-1}$~$a_{1g}$ and S 2p$^{-1}$~$t_{2g}$ states as well as a broad band attributed to the S 2p$^{-1}$~$e_{g}$ doublet, in agreement with reported synchrotron data \cite{HudsonPRA1993} and our B3LYP TDDFT calculations (orange sticks). Our discussion focuses now on the dynamics outside of the temporal overlap of the pulses, as opposed to the recent work of Rupprecht \textit{et al.} \cite{RupprechtPRL2022}. The transient absorption of SF$_6$ after excitation by the $\approx$~$6 \times 10^{14}$ W/cm$^2$ vis-NIR pulse is shown in Fig.~\ref{fig:Transient}b.  Clear oscillations of the absorbance are observed at positive delays for the five peaks with a period of $\approx$~43 fs. The oscillations are long-lived, and the Fourier transform in Fig.~\ref{fig:Transient}c reveals a single feature at 775~cm$^{-1}$ corresponding to the $\nu_1$ vibrational mode, visible at the energies of all the core-excited levels. With short pump pulses in this intensity range, previous work showed that the three Raman-active modes can be simultaneously excited \cite{wagnerPNAS2006,ferreJPB2014,ferreNatComm2015,BaykushevaPRL2016}, but only one is observed here. The positive features appearing at positive delays in-between the assigned peaks in Fig.~\ref{fig:Transient}b may be attributed to strong-field dissociation of SF$_6$, although they differ in shape from what was observed in \cite{pertotScience2017}. Bands attributed to the formation of SF$_5^+$ in Ref.~\cite{pertotScience2017} were not observed under our experimental conditions, despite our higher spectral resolution. 

\begin{figure}
    \centering
    \includegraphics[width=\linewidth]{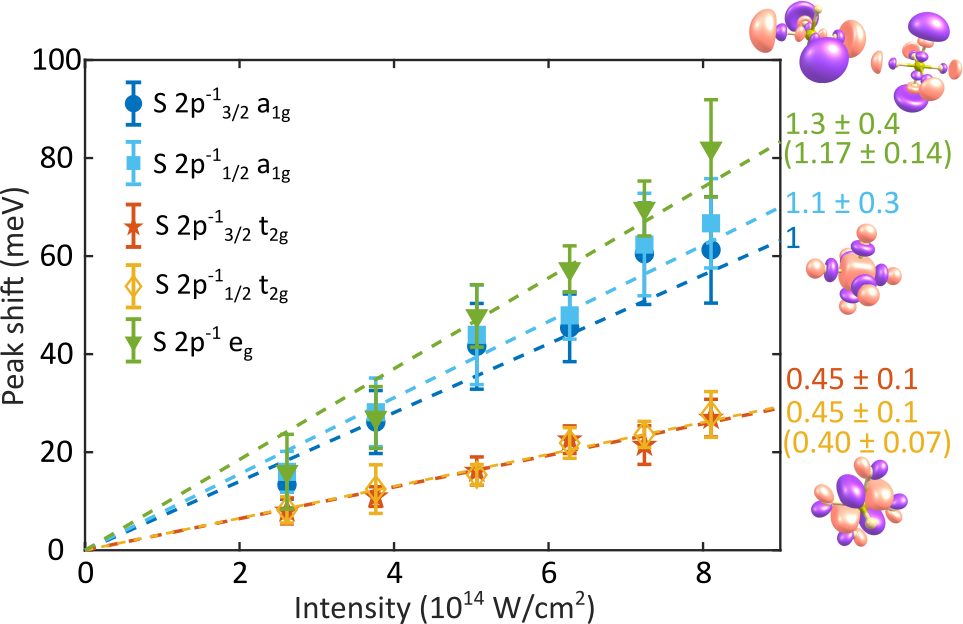}
    \caption{Amplitude of the absorption peak shift as a function of the pump intensity for the five S core-excited states and their linear fit (dashed line). The error bars represent the 95 \% confidence bounds. The relative slopes are indicated on the right, with numbers in parenthesis obtained from a fitting procedure including other data sets described in \cite{RossArXiv2022,SeeSM}. Illustrations of the $e_g$, $a_{1g}$ and $t_{2g}$ molecular orbitals are shown next to the corresponding state.}
    \label{fig:Powerscan}%
\end{figure}

Mapping the same vibrational coherence onto five different core-excited states allows us to compare them. To quantitatively characterize the PES gradients of the five states for the $\nu_1$ normal mode, we use the vis-NIR pump intensity to control the number of vibrational states included in the superposition and therefore the nuclear wavepacket. For ISRS excitation with an electric field $\mathcal{E}$, the nuclear wavefunction $\chi (q,t)$ satisfies the time-dependent Schrödinger equation (in atomic units):
\begin{eqnarray}\label{eq:shr}
    i \frac{\partial \chi (q,t)}{\partial t} &=& \left( -\frac{1}{2 \mu} \frac{\partial^2}{\partial q^2} + V(q)\right. \nonumber\\
    &-& \left. \frac{1}{2} \sum_{i,j} \alpha_{ij}(q) \mathcal{E}_i(t) \mathcal{E}_j(t) \right ) \chi(q,t)
\end{eqnarray}
where $\mu$ is the reduced mass of the normal mode, $V$ the ground state PES, $\alpha$ the polarizability tensor and $\mathcal{E}_i$ the amplitude of the electric field of the laser pulse along the $i$ axis. For the symmetric stretch mode $\nu_1$ the normal coordinate $q$ identifies with the S-F distance. Figure~\ref{fig:Wavepacket}a displays the squared nuclear wavepacket $|\chi(q,t)|^2$ in the ground state found from the numerical solution of Eq.(\ref{eq:shr}) with a $6 \times 10^{14}$  W/cm$^2$, 6 fs fwhm Gaussian pump pulse and $\alpha$ and $V(q)$ extracted from DFT B3LYP/def2-TZVP quantum-chemical simulations under constrained O$_h$ symmetry with the GAMESS software. The theoretical PES results in a smaller vibrational frequency than the tabulated value (728 cm$^{-1}$), giving a slightly longer period of the wavepacket oscillations in Fig. ~\ref{fig:Wavepacket}a compared to the experiment. At this pump intensity, vibrational levels up to $\nu_1 = 3$ are populated by ISRS \cite{SeeSM}, and the S-F distance changes by $\pm 2.4$~\%. As the center of the wavepacket oscillates, it is mapped onto the five core-excited PESs. The absorbance measured after excitation with the $6 \times 10^{14}$ W/cm$^2$ vis-NIR pulse with 3 fs delay steps is shown in Fig.~\ref{fig:Wavepacket}b-d. Apart from a slight difference in frequency originating from the calculated ground state potential, the absorbance oscillations nicely follow the nuclear wavepacket. At each delay, the absorption features are fitted by a Gaussian or Lorentzian function depending on the core-excited state \cite{HudsonPRA1993}. Their central energy oscillates as a function of the vis-NIR - SXR delay with a period of $T_1 = 43$~fs. The oscillations of the central energy are then fitted to a cosine function. Their amplitudes are different for the five core-excited states and inform on their relative PES gradients. The vis-NIR pump intensity is varied with a broadband combination of half-waveplate and polarizer between $2.6 \times 10^{14}$ and $8 \times 10^{14}$ W/cm$^2$. This allows us to incorporate greater or fewer vibrational states in the coherent superposition. Calculations show that the nuclear wavepacket spatial excursion is linear with the laser intensity in this range\cite{SeeSM}. A SXR transient absorption spectrogram is measured for six different intensities, all other parameters remaining identical. The amplitude of the central energy oscillation for the five core-excited states at each pump intensity is reported in Fig.~\ref{fig:Powerscan}. For the same nuclear geometry change in the ground state (i.e. at a given pump intensity), the SXR transition energies to the S~2p$^{-1}$~$a_{1g}$ and S~2p$^{-1}$~$e_{g}$ states have wider excursions from the equilibrium geometry transition compared to the S 2p$^{-1}$ $t_{2g}$ state. This result reveals the larger displacement of the PES along the S-F bond upon excitation to the S~2p$^{-1}$~$a_{1g}$ and S~2p$^{-1}$~$e_{g}$ states compared to S 2p$^{-1}$ $t_{2g}$. This different behavior reflects the non-bonding character of the $t_{2g}$ molecular orbital, whereas both the $a_{1g}$ and $e_g$ orbitals are anti-bonding along the S-F bonds (Fig.~\ref{fig:Powerscan}) therefore the electronic energy is more dependent on the internuclear distance. 

More quantitatively, the linear increases of the energy shifts with the pump intensity - which is the spatial extension of the nuclear wavefunction - can be extracted from a fit of Fig.~\ref{fig:Powerscan}. These shifts are directly related to the gradients of the PESs along the S-F distance. With the lowest core-excited state S~2p$^{-1}_{3/2}$~$a_{1g}$ taken as a reference, the relative PESs gradients are 1.1 $\pm$ 0.3 (S~2p$^{-1}_{1/2}$~$a_{1g}$), 0.45 $\pm$ 0.1 (S~2p$^{-1}_{3/2}$~$t_{2g}$), 0.45 $\pm$ 0.1 (S~2p$^{-1}_{1/2}$~$t_{2g}$), and 1.3 $\pm$ 0.4 (S~2p$^{-1}$~$e_{g}$) (Fig.~\ref{fig:Powerscan}). The results are consistent with the calculated PESs presented in Fig.~\ref{fig:Pot_v1}. As typically observed for core-excited states \cite{MarchenkoPRL2017}, these PESs have steep gradients along the symmetric stretch mode $\nu_1$, of the order of 10 to 30~eV/\AA. At the equilibrium geometry, the gradients of the calculated PESs relative to S~2p$^{-1}_{3/2}$~$a_{1g}$ are 1.01 (S~2p$^{-1}_{1/2}$~$a_{1g}$), 0.38 (S~2p$^{-1}_{3/2}$~$t_{2g}$), 0.38 (S~2p$^{-1}_{1/2}$~$t_{2g}$), 1.05 (S~2p$^{-1}_{3/2}$~2$e_{g}$), and 1.05 (S~2p$^{-1}_{1/2}$~2$e_{g}$), in good agreement with the experimental values. The spectral broadening caused by the large gradient and additional spectral background of multielectron character \cite{kivimakiJESRP2016} does not allow the resolution of the $e_g$ doublet in practice. The calculated PESs for the ground state and the two lowest core-excited states of $a_{1g}$ symmetry along the normal coordinates of the two other Raman-active modes ($\nu_2$ and $\nu_5$) are shown in Fig.~\ref{fig:Pot_v1}b. The core-excited and ground states are relatively parallel, confirming the lack of observed oscillations in the experiment. A similar behaviour is expected for the higher $e_g$ and $t_{2g}$ core-excited states, but vibrations along the non-totally symmetric $\nu_2$ and $\nu_5$ modes lift the degeneracy of these states, making the calculations of the PESs more complex and beyond the scope of this work.

Vibrational dynamics resulting from strong-field ionization have previously been observed with X-ray transient absorption spectroscopy \cite{HoslerPRA2013,WeiNatComm2017,saitoOptica2019,timmersNatComm2019,KobayashiPRA2020a,KobayashiPRA2020b,zinchenkoScience2021}. Here, the use of ISRS excitation provides a controlled vibrational wavepacket in the ground electronic state of the molecule. This in turn enables the characterization of multiple core-excited PESs. These results are an experimental demonstration of how to probe different regions of the core-excited PESs with the IR pump/X-ray probe scheme theoretically proposed over fifteen years ago \cite{FelicissimoPRA2005,GuimaraesPRA2005,felicissimoJCP2005}. Taking advantage of the element specificity of X-ray spectroscopy, multidimensional unexplored regions of core-excited PESs in many systems are accessible with this scheme, implemented either on table-top sources of femtosecond X-ray pulses or Free Electron Lasers. These unexplored regions are expected to drive nuclear dynamics, such as proton transfer in oxygen-core-excited water dimers \cite{felicissimoJCP2005}, and could therefore be used to control chemical reactions in short-lived core-excited states.\newline

The authors acknowledge fruitful discussions with Yann Mairesse and Marc Simon. This work is supported by the Gas Phase Chemical Physics program of the U.S. Department of Energy, Office of Science, Office of Basic Energy Sciences, Chemical Sciences, Geosciences and Biosciences Division (DE-AC02-05CH11231, FWP no. CHPHYS01 (D.M.N., S.R.L.)), the National Science Foundation (Grant No. CHE-1951317 and No. CHE-1660417 (S.R.L.), NSF MRI 1624322 for equipment and A.D.R.), the U.S. Army Research Office under Grant No. W911NF-14-1-0383 (D.M.N., S.R.L.). A.D.R. is also supported by the U.S. Army Research Office Grant No. W011NF-20-1-0127 (D.M.N.) and the W.M. Keck Foundation Grant No. 042982 (S.R.L.). V.K. acknowledges support from the Swedish Research Council (VR) project No. 2019-03470. P.K. and S.B. acknowledge Russian Foundation for Basic Research (RFBR), Project No. 19-29-12015. L.B. acknowledges support from the Miller Institute for Basic Research in Science at UC Berkeley.

\bibliography{Bibliography}

%apsrev4-2.bst 2019-01-14 (MD) hand-edited version of apsrev4-1.bst
%Control: key (0)
%Control: author (8) initials jnrlst
%Control: editor formatted (1) identically to author
%Control: production of article title (0) allowed
%Control: page (0) single
%Control: year (1) truncated
%Control: production of eprint (0) enabled
\begin{thebibliography}{49}%
\makeatletter
\providecommand \@ifxundefined [1]{%
 \@ifx{#1\undefined}
}%
\providecommand \@ifnum [1]{%
 \ifnum #1\expandafter \@firstoftwo
 \else \expandafter \@secondoftwo
 \fi
}%
\providecommand \@ifx [1]{%
 \ifx #1\expandafter \@firstoftwo
 \else \expandafter \@secondoftwo
 \fi
}%
\providecommand \natexlab [1]{#1}%
\providecommand \enquote  [1]{``#1''}%
\providecommand \bibnamefont  [1]{#1}%
\providecommand \bibfnamefont [1]{#1}%
\providecommand \citenamefont [1]{#1}%
\providecommand \href@noop [0]{\@secondoftwo}%
\providecommand \href [0]{\begingroup \@sanitize@url \@href}%
\providecommand \@href[1]{\@@startlink{#1}\@@href}%
\providecommand \@@href[1]{\endgroup#1\@@endlink}%
\providecommand \@sanitize@url [0]{\catcode `\\12\catcode `\$12\catcode
  `\&12\catcode `\#12\catcode `\^12\catcode `\_12\catcode `\%12\relax}%
\providecommand \@@startlink[1]{}%
\providecommand \@@endlink[0]{}%
\providecommand \url  [0]{\begingroup\@sanitize@url \@url }%
\providecommand \@url [1]{\endgroup\@href {#1}{\urlprefix }}%
\providecommand \urlprefix  [0]{URL }%
\providecommand \Eprint [0]{\href }%
\providecommand \doibase [0]{https://doi.org/}%
\providecommand \selectlanguage [0]{\@gobble}%
\providecommand \bibinfo  [0]{\@secondoftwo}%
\providecommand \bibfield  [0]{\@secondoftwo}%
\providecommand \translation [1]{[#1]}%
\providecommand \BibitemOpen [0]{}%
\providecommand \bibitemStop [0]{}%
\providecommand \bibitemNoStop [0]{.\EOS\space}%
\providecommand \EOS [0]{\spacefactor3000\relax}%
\providecommand \BibitemShut  [1]{\csname bibitem#1\endcsname}%
\let\auto@bib@innerbib\@empty
%</preamble>
\bibitem [{\citenamefont {Gelmukhanov}\ and\ \citenamefont
  {\AA{}gren}(1996)}]{GelmukhanovPRA1996}%
  \BibitemOpen
  \bibfield  {author} {\bibinfo {author} {\bibfnamefont {F.}~\bibnamefont
  {Gelmukhanov}}\ and\ \bibinfo {author} {\bibfnamefont {H.}~\bibnamefont
  {\AA{}gren}},\ }\bibfield  {title} {\bibinfo {title} {X-ray resonant
  scattering involving dissociative states},\ }\href
  {https://doi.org/10.1103/PhysRevA.54.379} {\bibfield  {journal} {\bibinfo
  {journal} {Phys. Rev. A}\ }\textbf {\bibinfo {volume} {54}},\ \bibinfo
  {pages} {379} (\bibinfo {year} {1996})}\BibitemShut {NoStop}%
\bibitem [{\citenamefont {Simon}\ \emph {et~al.}(1997)\citenamefont {Simon},
  \citenamefont {Miron}, \citenamefont {Leclercq}, \citenamefont {Morin},
  \citenamefont {Ueda}, \citenamefont {Sato}, \citenamefont {Tanaka},\ and\
  \citenamefont {Kayanuma}}]{SimonPRL1997}%
  \BibitemOpen
  \bibfield  {author} {\bibinfo {author} {\bibfnamefont {M.}~\bibnamefont
  {Simon}}, \bibinfo {author} {\bibfnamefont {C.}~\bibnamefont {Miron}},
  \bibinfo {author} {\bibfnamefont {N.}~\bibnamefont {Leclercq}}, \bibinfo
  {author} {\bibfnamefont {P.}~\bibnamefont {Morin}}, \bibinfo {author}
  {\bibfnamefont {K.}~\bibnamefont {Ueda}}, \bibinfo {author} {\bibfnamefont
  {Y.}~\bibnamefont {Sato}}, \bibinfo {author} {\bibfnamefont {S.}~\bibnamefont
  {Tanaka}},\ and\ \bibinfo {author} {\bibfnamefont {Y.}~\bibnamefont
  {Kayanuma}},\ }\bibfield  {title} {\bibinfo {title} {Nuclear motion of core
  excited {BF}$_{3}$ probed by high resolution resonant auger spectroscopy},\
  }\href {https://doi.org/10.1103/PhysRevLett.79.3857} {\bibfield  {journal}
  {\bibinfo  {journal} {Phys. Rev. Lett.}\ }\textbf {\bibinfo {volume} {79}},\
  \bibinfo {pages} {3857} (\bibinfo {year} {1997})}\BibitemShut {NoStop}%
\bibitem [{\citenamefont {Marchenko}\ \emph {et~al.}(2017)\citenamefont
  {Marchenko}, \citenamefont {Goldsztejn}, \citenamefont {J\"ank\"al\"a},
  \citenamefont {Travnikova}, \citenamefont {Journel}, \citenamefont
  {Guillemin}, \citenamefont {Sisourat}, \citenamefont {C\'eolin},
  \citenamefont {\ifmmode~\check{Z}\else \v{Z}\fi{}itnik}, \citenamefont
  {Kav\ifmmode \check{c}\else \v{c}\fi{}i\ifmmode~\check{c}\else \v{c}\fi{}},
  \citenamefont {Bu\ifmmode~\check{c}\else \v{c}\fi{}ar}, \citenamefont
  {Miheli\ifmmode~\check{c}\else \v{c}\fi{}}, \citenamefont {de~Miranda},
  \citenamefont {Ismail}, \citenamefont {Lago}, \citenamefont {Gel'mukhanov},
  \citenamefont {P\"uttner}, \citenamefont {Piancastelli},\ and\ \citenamefont
  {Simon}}]{MarchenkoPRL2017}%
  \BibitemOpen
  \bibfield  {author} {\bibinfo {author} {\bibfnamefont {T.}~\bibnamefont
  {Marchenko}}, \bibinfo {author} {\bibfnamefont {G.}~\bibnamefont
  {Goldsztejn}}, \bibinfo {author} {\bibfnamefont {K.}~\bibnamefont
  {J\"ank\"al\"a}}, \bibinfo {author} {\bibfnamefont {O.}~\bibnamefont
  {Travnikova}}, \bibinfo {author} {\bibfnamefont {L.}~\bibnamefont {Journel}},
  \bibinfo {author} {\bibfnamefont {R.}~\bibnamefont {Guillemin}}, \bibinfo
  {author} {\bibfnamefont {N.}~\bibnamefont {Sisourat}}, \bibinfo {author}
  {\bibfnamefont {D.}~\bibnamefont {C\'eolin}}, \bibinfo {author}
  {\bibfnamefont {M.}~\bibnamefont {\ifmmode~\check{Z}\else \v{Z}\fi{}itnik}},
  \bibinfo {author} {\bibfnamefont {M.}~\bibnamefont {Kav\ifmmode
  \check{c}\else \v{c}\fi{}i\ifmmode~\check{c}\else \v{c}\fi{}}}, \bibinfo
  {author} {\bibfnamefont {K.}~\bibnamefont {Bu\ifmmode~\check{c}\else
  \v{c}\fi{}ar}}, \bibinfo {author} {\bibfnamefont {A.}~\bibnamefont
  {Miheli\ifmmode~\check{c}\else \v{c}\fi{}}}, \bibinfo {author} {\bibfnamefont
  {B.~C.}\ \bibnamefont {de~Miranda}}, \bibinfo {author} {\bibfnamefont
  {I.}~\bibnamefont {Ismail}}, \bibinfo {author} {\bibfnamefont {A.~F.}\
  \bibnamefont {Lago}}, \bibinfo {author} {\bibfnamefont {F.}~\bibnamefont
  {Gel'mukhanov}}, \bibinfo {author} {\bibfnamefont {R.}~\bibnamefont
  {P\"uttner}}, \bibinfo {author} {\bibfnamefont {M.~N.}\ \bibnamefont
  {Piancastelli}},\ and\ \bibinfo {author} {\bibfnamefont {M.}~\bibnamefont
  {Simon}},\ }\bibfield  {title} {\bibinfo {title} {Potential energy surface
  reconstruction and lifetime determination of molecular double-core-hole
  states in the hard x-ray regime},\ }\href
  {https://doi.org/10.1103/PhysRevLett.119.133001} {\bibfield  {journal}
  {\bibinfo  {journal} {Phys. Rev. Lett.}\ }\textbf {\bibinfo {volume} {119}},\
  \bibinfo {pages} {133001} (\bibinfo {year} {2017})}\BibitemShut {NoStop}%
\bibitem [{\citenamefont {Felic\'{\i}ssimo}\ \emph {et~al.}(2005)\citenamefont
  {Felic\'{\i}ssimo}, \citenamefont {Guimar\~aes},\ and\ \citenamefont
  {Gel'mukhanov}}]{FelicissimoPRA2005}%
  \BibitemOpen
  \bibfield  {author} {\bibinfo {author} {\bibfnamefont {V.~C.}\ \bibnamefont
  {Felic\'{\i}ssimo}}, \bibinfo {author} {\bibfnamefont {F.~F.}\ \bibnamefont
  {Guimar\~aes}},\ and\ \bibinfo {author} {\bibfnamefont {F.}~\bibnamefont
  {Gel'mukhanov}},\ }\bibfield  {title} {\bibinfo {title} {Enhancement of the
  recoil effect in {X}-ray photoelectron spectra of molecules driven by a
  strong {IR} field},\ }\href {https://doi.org/10.1103/PhysRevA.72.023414}
  {\bibfield  {journal} {\bibinfo  {journal} {Phys. Rev. A}\ }\textbf {\bibinfo
  {volume} {72}},\ \bibinfo {pages} {023414} (\bibinfo {year}
  {2005})}\BibitemShut {NoStop}%
\bibitem [{\citenamefont {Guimar\~aes}\ \emph {et~al.}(2005)\citenamefont
  {Guimar\~aes}, \citenamefont {Kimberg}, \citenamefont {Felic\'{\i}ssimo},
  \citenamefont {Gel'mukhanov}, \citenamefont {Cesar},\ and\ \citenamefont
  {\AA{}gren}}]{GuimaraesPRA2005}%
  \BibitemOpen
  \bibfield  {author} {\bibinfo {author} {\bibfnamefont {F.~F.}\ \bibnamefont
  {Guimar\~aes}}, \bibinfo {author} {\bibfnamefont {V.}~\bibnamefont
  {Kimberg}}, \bibinfo {author} {\bibfnamefont {V.~C.}\ \bibnamefont
  {Felic\'{\i}ssimo}}, \bibinfo {author} {\bibfnamefont {F.}~\bibnamefont
  {Gel'mukhanov}}, \bibinfo {author} {\bibfnamefont {A.}~\bibnamefont
  {Cesar}},\ and\ \bibinfo {author} {\bibfnamefont {H.}~\bibnamefont
  {\AA{}gren}},\ }\bibfield  {title} {\bibinfo {title} {Infrared--x-ray
  pump-probe spectroscopy of the {NO} molecule},\ }\href
  {https://doi.org/10.1103/PhysRevA.72.012714} {\bibfield  {journal} {\bibinfo
  {journal} {Phys. Rev. A}\ }\textbf {\bibinfo {volume} {72}},\ \bibinfo
  {pages} {012714} (\bibinfo {year} {2005})}\BibitemShut {NoStop}%
\bibitem [{\citenamefont {Felicissimo}\ \emph {et~al.}(2005)\citenamefont
  {Felicissimo}, \citenamefont {Guimaraes}, \citenamefont {Gel’mukhanov},
  \citenamefont {Cesar},\ and\ \citenamefont {{\AA}gren}}]{felicissimoJCP2005}%
  \BibitemOpen
  \bibfield  {author} {\bibinfo {author} {\bibfnamefont {V.~C.}\ \bibnamefont
  {Felicissimo}}, \bibinfo {author} {\bibfnamefont {F.~F.}\ \bibnamefont
  {Guimaraes}}, \bibinfo {author} {\bibfnamefont {F.}~\bibnamefont
  {Gel’mukhanov}}, \bibinfo {author} {\bibfnamefont {A.}~\bibnamefont
  {Cesar}},\ and\ \bibinfo {author} {\bibfnamefont {H.}~\bibnamefont
  {{\AA}gren}},\ }\bibfield  {title} {\bibinfo {title} {The principles of
  infrared-{X}-ray pump-probe spectroscopy. applications on proton transfer in
  core-ionized water dimers},\ }\href {https://doi.org/10.1063/1.1860312}
  {\bibfield  {journal} {\bibinfo  {journal} {The Journal of chemical physics}\
  }\textbf {\bibinfo {volume} {122}},\ \bibinfo {pages} {094319} (\bibinfo
  {year} {2005})}\BibitemShut {NoStop}%
\bibitem [{\citenamefont {Carniato}\ \emph {et~al.}(2007)\citenamefont
  {Carniato}, \citenamefont {Ta{\"\i}eb}, \citenamefont {Guillemin},
  \citenamefont {Journel}, \citenamefont {Simon},\ and\ \citenamefont
  {Gel’mukhanov}}]{carniatoCPL2007}%
  \BibitemOpen
  \bibfield  {author} {\bibinfo {author} {\bibfnamefont {S.}~\bibnamefont
  {Carniato}}, \bibinfo {author} {\bibfnamefont {R.}~\bibnamefont
  {Ta{\"\i}eb}}, \bibinfo {author} {\bibfnamefont {R.}~\bibnamefont
  {Guillemin}}, \bibinfo {author} {\bibfnamefont {L.}~\bibnamefont {Journel}},
  \bibinfo {author} {\bibfnamefont {M.}~\bibnamefont {Simon}},\ and\ \bibinfo
  {author} {\bibfnamefont {F.}~\bibnamefont {Gel’mukhanov}},\ }\bibfield
  {title} {\bibinfo {title} {{K}-{L} resonant {X}-ray {Raman} scattering as a
  tool for potential energy surface mapping},\ }\href
  {https://doi.org/https://doi.org/10.1016/j.cplett.2007.03.100} {\bibfield
  {journal} {\bibinfo  {journal} {Chemical physics letters}\ }\textbf {\bibinfo
  {volume} {439}},\ \bibinfo {pages} {402} (\bibinfo {year}
  {2007})}\BibitemShut {NoStop}%
\bibitem [{\citenamefont {Engin}\ \emph {et~al.}(2012)\citenamefont {Engin},
  \citenamefont {Sisourat}, \citenamefont {Selles}, \citenamefont
  {Ta{\"\i}eb},\ and\ \citenamefont {Carniato}}]{enginCPL2012}%
  \BibitemOpen
  \bibfield  {author} {\bibinfo {author} {\bibfnamefont {S.}~\bibnamefont
  {Engin}}, \bibinfo {author} {\bibfnamefont {N.}~\bibnamefont {Sisourat}},
  \bibinfo {author} {\bibfnamefont {P.}~\bibnamefont {Selles}}, \bibinfo
  {author} {\bibfnamefont {R.}~\bibnamefont {Ta{\"\i}eb}},\ and\ \bibinfo
  {author} {\bibfnamefont {S.}~\bibnamefont {Carniato}},\ }\bibfield  {title}
  {\bibinfo {title} {Probing {IR}-{Raman} vibrationally excited molecules with
  {X}-ray spectroscopy},\ }\href
  {https://doi.org/https://doi.org/10.1016/j.cplett.2012.03.062} {\bibfield
  {journal} {\bibinfo  {journal} {Chemical Physics Letters}\ }\textbf {\bibinfo
  {volume} {535}},\ \bibinfo {pages} {192} (\bibinfo {year}
  {2012})}\BibitemShut {NoStop}%
\bibitem [{\citenamefont {Ignatova}\ \emph {et~al.}(2017)\citenamefont
  {Ignatova}, \citenamefont {da~Cruz}, \citenamefont {Couto}, \citenamefont
  {Ertan}, \citenamefont {Odelius}, \citenamefont {\AA{}gren}, \citenamefont
  {Guimar\~aes}, \citenamefont {Zimin}, \citenamefont {Polyutov}, \citenamefont
  {Gel'mukhanov},\ and\ \citenamefont {Kimberg}}]{IgnatovaPRA2017}%
  \BibitemOpen
  \bibfield  {author} {\bibinfo {author} {\bibfnamefont {N.}~\bibnamefont
  {Ignatova}}, \bibinfo {author} {\bibfnamefont {V.~V.}\ \bibnamefont
  {da~Cruz}}, \bibinfo {author} {\bibfnamefont {R.~C.}\ \bibnamefont {Couto}},
  \bibinfo {author} {\bibfnamefont {E.}~\bibnamefont {Ertan}}, \bibinfo
  {author} {\bibfnamefont {M.}~\bibnamefont {Odelius}}, \bibinfo {author}
  {\bibfnamefont {H.}~\bibnamefont {\AA{}gren}}, \bibinfo {author}
  {\bibfnamefont {F.~F.}\ \bibnamefont {Guimar\~aes}}, \bibinfo {author}
  {\bibfnamefont {A.}~\bibnamefont {Zimin}}, \bibinfo {author} {\bibfnamefont
  {S.~P.}\ \bibnamefont {Polyutov}}, \bibinfo {author} {\bibfnamefont
  {F.}~\bibnamefont {Gel'mukhanov}},\ and\ \bibinfo {author} {\bibfnamefont
  {V.}~\bibnamefont {Kimberg}},\ }\bibfield  {title} {\bibinfo {title}
  {Infrared-pump--x-ray-probe spectroscopy of vibrationally excited
  molecules},\ }\href {https://doi.org/10.1103/PhysRevA.95.042502} {\bibfield
  {journal} {\bibinfo  {journal} {Phys. Rev. A}\ }\textbf {\bibinfo {volume}
  {95}},\ \bibinfo {pages} {042502} (\bibinfo {year} {2017})}\BibitemShut
  {NoStop}%
\bibitem [{\citenamefont {Young}\ \emph {et~al.}(2018)\citenamefont {Young},
  \citenamefont {Ueda}, \citenamefont {G{\"u}hr}, \citenamefont {Bucksbaum},
  \citenamefont {Simon}, \citenamefont {Mukamel}, \citenamefont {Rohringer},
  \citenamefont {Prince}, \citenamefont {Masciovecchio}, \citenamefont {Meyer}
  \emph {et~al.}}]{YoungJPhysB2018}%
  \BibitemOpen
  \bibfield  {author} {\bibinfo {author} {\bibfnamefont {L.}~\bibnamefont
  {Young}}, \bibinfo {author} {\bibfnamefont {K.}~\bibnamefont {Ueda}},
  \bibinfo {author} {\bibfnamefont {M.}~\bibnamefont {G{\"u}hr}}, \bibinfo
  {author} {\bibfnamefont {P.~H.}\ \bibnamefont {Bucksbaum}}, \bibinfo {author}
  {\bibfnamefont {M.}~\bibnamefont {Simon}}, \bibinfo {author} {\bibfnamefont
  {S.}~\bibnamefont {Mukamel}}, \bibinfo {author} {\bibfnamefont
  {N.}~\bibnamefont {Rohringer}}, \bibinfo {author} {\bibfnamefont {K.~C.}\
  \bibnamefont {Prince}}, \bibinfo {author} {\bibfnamefont {C.}~\bibnamefont
  {Masciovecchio}}, \bibinfo {author} {\bibfnamefont {M.}~\bibnamefont
  {Meyer}}, \emph {et~al.},\ }\bibfield  {title} {\bibinfo {title} {Roadmap of
  ultrafast {X}-ray atomic and molecular physics},\ }\href
  {https://doi.org/10.1088/1361-6455/aa9735} {\bibfield  {journal} {\bibinfo
  {journal} {Journal of Physics B: Atomic, Molecular and Optical Physics}\
  }\textbf {\bibinfo {volume} {51}},\ \bibinfo {pages} {032003} (\bibinfo
  {year} {2018})}\BibitemShut {NoStop}%
\bibitem [{\citenamefont {Geneaux}\ \emph {et~al.}(2019)\citenamefont
  {Geneaux}, \citenamefont {Marroux}, \citenamefont {Guggenmos}, \citenamefont
  {Neumark},\ and\ \citenamefont {Leone}}]{geneauxPhilTrans2019}%
  \BibitemOpen
  \bibfield  {author} {\bibinfo {author} {\bibfnamefont {R.}~\bibnamefont
  {Geneaux}}, \bibinfo {author} {\bibfnamefont {H.~J.}\ \bibnamefont
  {Marroux}}, \bibinfo {author} {\bibfnamefont {A.}~\bibnamefont {Guggenmos}},
  \bibinfo {author} {\bibfnamefont {D.~M.}\ \bibnamefont {Neumark}},\ and\
  \bibinfo {author} {\bibfnamefont {S.~R.}\ \bibnamefont {Leone}},\ }\bibfield
  {title} {\bibinfo {title} {Transient absorption spectroscopy using high
  harmonic generation: a review of ultrafast {X}-ray dynamics in molecules and
  solids},\ }\href@noop {} {\bibfield  {journal} {\bibinfo  {journal}
  {Philosophical Transactions of the Royal Society A}\ }\textbf {\bibinfo
  {volume} {377}},\ \bibinfo {pages} {20170463} (\bibinfo {year}
  {2019})}\BibitemShut {NoStop}%
\bibitem [{\citenamefont {Hosler}\ and\ \citenamefont
  {Leone}(2013)}]{HoslerPRA2013}%
  \BibitemOpen
  \bibfield  {author} {\bibinfo {author} {\bibfnamefont {E.~R.}\ \bibnamefont
  {Hosler}}\ and\ \bibinfo {author} {\bibfnamefont {S.~R.}\ \bibnamefont
  {Leone}},\ }\bibfield  {title} {\bibinfo {title} {Characterization of
  vibrational wave packets by core-level high-harmonic transient absorption
  spectroscopy},\ }\href {https://doi.org/10.1103/PhysRevA.88.023420}
  {\bibfield  {journal} {\bibinfo  {journal} {Phys. Rev. A}\ }\textbf {\bibinfo
  {volume} {88}},\ \bibinfo {pages} {023420} (\bibinfo {year}
  {2013})}\BibitemShut {NoStop}%
\bibitem [{\citenamefont {Kobayashi}\ \emph
  {et~al.}(2020{\natexlab{a}})\citenamefont {Kobayashi}, \citenamefont {Chang},
  \citenamefont {Poullain}, \citenamefont {Scutelnic}, \citenamefont {Zeng},
  \citenamefont {Neumark},\ and\ \citenamefont {Leone}}]{KobayashiPRA2020a}%
  \BibitemOpen
  \bibfield  {author} {\bibinfo {author} {\bibfnamefont {Y.}~\bibnamefont
  {Kobayashi}}, \bibinfo {author} {\bibfnamefont {K.~F.}\ \bibnamefont
  {Chang}}, \bibinfo {author} {\bibfnamefont {S.~M.}\ \bibnamefont {Poullain}},
  \bibinfo {author} {\bibfnamefont {V.}~\bibnamefont {Scutelnic}}, \bibinfo
  {author} {\bibfnamefont {T.}~\bibnamefont {Zeng}}, \bibinfo {author}
  {\bibfnamefont {D.~M.}\ \bibnamefont {Neumark}},\ and\ \bibinfo {author}
  {\bibfnamefont {S.~R.}\ \bibnamefont {Leone}},\ }\bibfield  {title} {\bibinfo
  {title} {Coherent electronic-vibrational dynamics in deuterium bromide probed
  via attosecond transient-absorption spectroscopy},\ }\href
  {https://doi.org/10.1103/PhysRevA.101.063414} {\bibfield  {journal} {\bibinfo
   {journal} {Phys. Rev. A}\ }\textbf {\bibinfo {volume} {101}},\ \bibinfo
  {pages} {063414} (\bibinfo {year} {2020}{\natexlab{a}})}\BibitemShut
  {NoStop}%
\bibitem [{\citenamefont {Kobayashi}\ \emph
  {et~al.}(2020{\natexlab{b}})\citenamefont {Kobayashi}, \citenamefont
  {Neumark},\ and\ \citenamefont {Leone}}]{KobayashiPRA2020b}%
  \BibitemOpen
  \bibfield  {author} {\bibinfo {author} {\bibfnamefont {Y.}~\bibnamefont
  {Kobayashi}}, \bibinfo {author} {\bibfnamefont {D.~M.}\ \bibnamefont
  {Neumark}},\ and\ \bibinfo {author} {\bibfnamefont {S.~R.}\ \bibnamefont
  {Leone}},\ }\bibfield  {title} {\bibinfo {title} {Attosecond {XUV} probing of
  vibronic quantum superpositions in {Br}$_{2}^{+}$},\ }\href
  {https://doi.org/10.1103/PhysRevA.102.051102} {\bibfield  {journal} {\bibinfo
   {journal} {Phys. Rev. A}\ }\textbf {\bibinfo {volume} {102}},\ \bibinfo
  {pages} {051102} (\bibinfo {year} {2020}{\natexlab{b}})}\BibitemShut
  {NoStop}%
\bibitem [{\citenamefont {Saito}\ \emph {et~al.}(2019)\citenamefont {Saito},
  \citenamefont {Sannohe}, \citenamefont {Ishii}, \citenamefont {Kanai},
  \citenamefont {Kosugi}, \citenamefont {Wu}, \citenamefont {Chew},
  \citenamefont {Han}, \citenamefont {Chang},\ and\ \citenamefont
  {Itatani}}]{saitoOptica2019}%
  \BibitemOpen
  \bibfield  {author} {\bibinfo {author} {\bibfnamefont {N.}~\bibnamefont
  {Saito}}, \bibinfo {author} {\bibfnamefont {H.}~\bibnamefont {Sannohe}},
  \bibinfo {author} {\bibfnamefont {N.}~\bibnamefont {Ishii}}, \bibinfo
  {author} {\bibfnamefont {T.}~\bibnamefont {Kanai}}, \bibinfo {author}
  {\bibfnamefont {N.}~\bibnamefont {Kosugi}}, \bibinfo {author} {\bibfnamefont
  {Y.}~\bibnamefont {Wu}}, \bibinfo {author} {\bibfnamefont {A.}~\bibnamefont
  {Chew}}, \bibinfo {author} {\bibfnamefont {S.}~\bibnamefont {Han}}, \bibinfo
  {author} {\bibfnamefont {Z.}~\bibnamefont {Chang}},\ and\ \bibinfo {author}
  {\bibfnamefont {J.}~\bibnamefont {Itatani}},\ }\bibfield  {title} {\bibinfo
  {title} {Real-time observation of electronic, vibrational, and rotational
  dynamics in nitric oxide with attosecond soft {X}-ray pulses at 400 {eV}},\
  }\href {https://doi.org/https://doi.org/10.1364/OPTICA.6.001542} {\bibfield
  {journal} {\bibinfo  {journal} {Optica}\ }\textbf {\bibinfo {volume} {6}},\
  \bibinfo {pages} {1542} (\bibinfo {year} {2019})}\BibitemShut {NoStop}%
\bibitem [{\citenamefont {Wei}\ \emph {et~al.}(2017)\citenamefont {Wei},
  \citenamefont {Li}, \citenamefont {Wang}, \citenamefont {See}, \citenamefont
  {Jhon}, \citenamefont {Zhang}, \citenamefont {Shi}, \citenamefont {Yang},\
  and\ \citenamefont {Loh}}]{WeiNatComm2017}%
  \BibitemOpen
  \bibfield  {author} {\bibinfo {author} {\bibfnamefont {Z.}~\bibnamefont
  {Wei}}, \bibinfo {author} {\bibfnamefont {J.}~\bibnamefont {Li}}, \bibinfo
  {author} {\bibfnamefont {L.}~\bibnamefont {Wang}}, \bibinfo {author}
  {\bibfnamefont {S.~T.}\ \bibnamefont {See}}, \bibinfo {author} {\bibfnamefont
  {M.~H.}\ \bibnamefont {Jhon}}, \bibinfo {author} {\bibfnamefont
  {Y.}~\bibnamefont {Zhang}}, \bibinfo {author} {\bibfnamefont
  {F.}~\bibnamefont {Shi}}, \bibinfo {author} {\bibfnamefont {M.}~\bibnamefont
  {Yang}},\ and\ \bibinfo {author} {\bibfnamefont {Z.-H.}\ \bibnamefont
  {Loh}},\ }\bibfield  {title} {\bibinfo {title} {Elucidating the origins of
  multimode vibrational coherences of polyatomic molecules induced by intense
  laser fields},\ }\href {https://doi.org/10.1038/s41467-017-00848-2}
  {\bibfield  {journal} {\bibinfo  {journal} {Nature communications}\ }\textbf
  {\bibinfo {volume} {8}},\ \bibinfo {pages} {1} (\bibinfo {year}
  {2017})}\BibitemShut {NoStop}%
\bibitem [{\citenamefont {Timmers}\ \emph {et~al.}(2019)\citenamefont
  {Timmers}, \citenamefont {Zhu}, \citenamefont {Li}, \citenamefont
  {Kobayashi}, \citenamefont {Sabbar}, \citenamefont {Hollstein}, \citenamefont
  {Reduzzi}, \citenamefont {Mart{\'\i}nez}, \citenamefont {Neumark},\ and\
  \citenamefont {Leone}}]{timmersNatComm2019}%
  \BibitemOpen
  \bibfield  {author} {\bibinfo {author} {\bibfnamefont {H.}~\bibnamefont
  {Timmers}}, \bibinfo {author} {\bibfnamefont {X.}~\bibnamefont {Zhu}},
  \bibinfo {author} {\bibfnamefont {Z.}~\bibnamefont {Li}}, \bibinfo {author}
  {\bibfnamefont {Y.}~\bibnamefont {Kobayashi}}, \bibinfo {author}
  {\bibfnamefont {M.}~\bibnamefont {Sabbar}}, \bibinfo {author} {\bibfnamefont
  {M.}~\bibnamefont {Hollstein}}, \bibinfo {author} {\bibfnamefont
  {M.}~\bibnamefont {Reduzzi}}, \bibinfo {author} {\bibfnamefont {T.~J.}\
  \bibnamefont {Mart{\'\i}nez}}, \bibinfo {author} {\bibfnamefont {D.~M.}\
  \bibnamefont {Neumark}},\ and\ \bibinfo {author} {\bibfnamefont {S.~R.}\
  \bibnamefont {Leone}},\ }\bibfield  {title} {\bibinfo {title} {Disentangling
  conical intersection and coherent molecular dynamics in methyl bromide with
  attosecond transient absorption spectroscopy},\ }\href
  {https://doi.org/https://doi.org/10.1038/s41467-019-10789-7} {\bibfield
  {journal} {\bibinfo  {journal} {Nature communications}\ }\textbf {\bibinfo
  {volume} {10}},\ \bibinfo {pages} {1} (\bibinfo {year} {2019})}\BibitemShut
  {NoStop}%
\bibitem [{\citenamefont {Zinchenko}\ \emph {et~al.}(2021)\citenamefont
  {Zinchenko}, \citenamefont {Ardana-Lamas}, \citenamefont {Seidu},
  \citenamefont {Neville}, \citenamefont {van~der Veen}, \citenamefont
  {Lanfaloni}, \citenamefont {Schuurman},\ and\ \citenamefont
  {W{\"o}rner}}]{zinchenkoScience2021}%
  \BibitemOpen
  \bibfield  {author} {\bibinfo {author} {\bibfnamefont {K.~S.}\ \bibnamefont
  {Zinchenko}}, \bibinfo {author} {\bibfnamefont {F.}~\bibnamefont
  {Ardana-Lamas}}, \bibinfo {author} {\bibfnamefont {I.}~\bibnamefont {Seidu}},
  \bibinfo {author} {\bibfnamefont {S.~P.}\ \bibnamefont {Neville}}, \bibinfo
  {author} {\bibfnamefont {J.}~\bibnamefont {van~der Veen}}, \bibinfo {author}
  {\bibfnamefont {V.~U.}\ \bibnamefont {Lanfaloni}}, \bibinfo {author}
  {\bibfnamefont {M.~S.}\ \bibnamefont {Schuurman}},\ and\ \bibinfo {author}
  {\bibfnamefont {H.~J.}\ \bibnamefont {W{\"o}rner}},\ }\bibfield  {title}
  {\bibinfo {title} {Sub-7-femtosecond conical-intersection dynamics probed at
  the carbon k-edge},\ }\href {https://doi.org/10.1126/science.abf1656}
  {\bibfield  {journal} {\bibinfo  {journal} {Science}\ }\textbf {\bibinfo
  {volume} {371}},\ \bibinfo {pages} {489} (\bibinfo {year}
  {2021})}\BibitemShut {NoStop}%
\bibitem [{\citenamefont {Poullain}\ \emph {et~al.}(2021)\citenamefont
  {Poullain}, \citenamefont {Kobayashi}, \citenamefont {Chang},\ and\
  \citenamefont {Leone}}]{PoullainPRA2021}%
  \BibitemOpen
  \bibfield  {author} {\bibinfo {author} {\bibfnamefont {S.~M.}\ \bibnamefont
  {Poullain}}, \bibinfo {author} {\bibfnamefont {Y.}~\bibnamefont {Kobayashi}},
  \bibinfo {author} {\bibfnamefont {K.~F.}\ \bibnamefont {Chang}},\ and\
  \bibinfo {author} {\bibfnamefont {S.~R.}\ \bibnamefont {Leone}},\ }\bibfield
  {title} {\bibinfo {title} {Visualizing coherent vibrational motion in the
  molecular iodine
  $b\phantom{\rule{0.28em}{0ex}}{}^{3}{\mathrm{\ensuremath{\Pi}}}_{{{0}^{+}}_{u}}$
  state using ultrafast xuv transient-absorption spectroscopy},\ }\href
  {https://doi.org/10.1103/PhysRevA.104.022817} {\bibfield  {journal} {\bibinfo
   {journal} {Phys. Rev. A}\ }\textbf {\bibinfo {volume} {104}},\ \bibinfo
  {pages} {022817} (\bibinfo {year} {2021})}\BibitemShut {NoStop}%
\bibitem [{\citenamefont {Chang}\ \emph {et~al.}(2022)\citenamefont {Chang},
  \citenamefont {Wang}, \citenamefont {Poullain}, \citenamefont
  {González-Vázquez}, \citenamefont {Bañares}, \citenamefont {Prendergast},
  \citenamefont {Neumark},\ and\ \citenamefont {Leone}}]{ChangJCP2022}%
  \BibitemOpen
  \bibfield  {author} {\bibinfo {author} {\bibfnamefont {K.~F.}\ \bibnamefont
  {Chang}}, \bibinfo {author} {\bibfnamefont {H.}~\bibnamefont {Wang}},
  \bibinfo {author} {\bibfnamefont {S.~M.}\ \bibnamefont {Poullain}}, \bibinfo
  {author} {\bibfnamefont {J.}~\bibnamefont {González-Vázquez}}, \bibinfo
  {author} {\bibfnamefont {L.}~\bibnamefont {Bañares}}, \bibinfo {author}
  {\bibfnamefont {D.}~\bibnamefont {Prendergast}}, \bibinfo {author}
  {\bibfnamefont {D.~M.}\ \bibnamefont {Neumark}},\ and\ \bibinfo {author}
  {\bibfnamefont {S.~R.}\ \bibnamefont {Leone}},\ }\bibfield  {title} {\bibinfo
  {title} {Conical intersection and coherent vibrational dynamics in alkyl
  iodides captured by attosecond transient absorption spectroscopy},\ }\href
  {https://doi.org/10.1063/5.0086775} {\bibfield  {journal} {\bibinfo
  {journal} {The Journal of Chemical Physics}\ }\textbf {\bibinfo {volume}
  {156}},\ \bibinfo {pages} {114304} (\bibinfo {year} {2022})}\BibitemShut
  {NoStop}%
\bibitem [{\citenamefont {Yan}\ \emph {et~al.}(1985)\citenamefont {Yan},
  \citenamefont {Gamble~Jr},\ and\ \citenamefont {Nelson}}]{yanJCP1985}%
  \BibitemOpen
  \bibfield  {author} {\bibinfo {author} {\bibfnamefont {Y.-X.}\ \bibnamefont
  {Yan}}, \bibinfo {author} {\bibfnamefont {E.~B.}\ \bibnamefont {Gamble~Jr}},\
  and\ \bibinfo {author} {\bibfnamefont {K.~A.}\ \bibnamefont {Nelson}},\
  }\bibfield  {title} {\bibinfo {title} {Impulsive stimulated scattering:
  General importance in femtosecond laser pulse interactions with matter, and
  spectroscopic applications},\ }\href {https://doi.org/10.1063/1.449708}
  {\bibfield  {journal} {\bibinfo  {journal} {The Journal of chemical physics}\
  }\textbf {\bibinfo {volume} {83}},\ \bibinfo {pages} {5391} (\bibinfo {year}
  {1985})}\BibitemShut {NoStop}%
\bibitem [{\citenamefont {Yan}\ and\ \citenamefont
  {Nelson}(1987)}]{yanJCP1987}%
  \BibitemOpen
  \bibfield  {author} {\bibinfo {author} {\bibfnamefont {Y.-X.}\ \bibnamefont
  {Yan}}\ and\ \bibinfo {author} {\bibfnamefont {K.~A.}\ \bibnamefont
  {Nelson}},\ }\bibfield  {title} {\bibinfo {title} {Impulsive stimulated light
  scattering. i. general theory},\ }\href {https://doi.org/10.1063/1.453733}
  {\bibfield  {journal} {\bibinfo  {journal} {The Journal of chemical physics}\
  }\textbf {\bibinfo {volume} {87}},\ \bibinfo {pages} {6240} (\bibinfo {year}
  {1987})}\BibitemShut {NoStop}%
\bibitem [{\citenamefont {Gerber}\ \emph {et~al.}(2017)\citenamefont {Gerber},
  \citenamefont {Yang}, \citenamefont {Zhu}, \citenamefont {Soifer},
  \citenamefont {Sobota}, \citenamefont {Rebec}, \citenamefont {Lee},
  \citenamefont {Jia}, \citenamefont {Moritz}, \citenamefont {Jia} \emph
  {et~al.}}]{gerberScience2017}%
  \BibitemOpen
  \bibfield  {author} {\bibinfo {author} {\bibfnamefont {S.}~\bibnamefont
  {Gerber}}, \bibinfo {author} {\bibfnamefont {S.-L.}\ \bibnamefont {Yang}},
  \bibinfo {author} {\bibfnamefont {D.}~\bibnamefont {Zhu}}, \bibinfo {author}
  {\bibfnamefont {H.}~\bibnamefont {Soifer}}, \bibinfo {author} {\bibfnamefont
  {J.}~\bibnamefont {Sobota}}, \bibinfo {author} {\bibfnamefont
  {S.}~\bibnamefont {Rebec}}, \bibinfo {author} {\bibfnamefont
  {J.}~\bibnamefont {Lee}}, \bibinfo {author} {\bibfnamefont {T.}~\bibnamefont
  {Jia}}, \bibinfo {author} {\bibfnamefont {B.}~\bibnamefont {Moritz}},
  \bibinfo {author} {\bibfnamefont {C.}~\bibnamefont {Jia}}, \emph {et~al.},\
  }\bibfield  {title} {\bibinfo {title} {Femtosecond electron-phonon lock-in by
  photoemission and {X}-ray free-electron laser},\ }\href
  {https://doi.org/10.1126/science.aak9946} {\bibfield  {journal} {\bibinfo
  {journal} {Science}\ }\textbf {\bibinfo {volume} {357}},\ \bibinfo {pages}
  {71} (\bibinfo {year} {2017})}\BibitemShut {NoStop}%
\bibitem [{\citenamefont {Hudson}\ \emph {et~al.}(1993)\citenamefont {Hudson},
  \citenamefont {Shirley}, \citenamefont {Domke}, \citenamefont {Remmers},
  \citenamefont {Puschmann}, \citenamefont {Mandel}, \citenamefont {Xue},\ and\
  \citenamefont {Kaindl}}]{HudsonPRA1993}%
  \BibitemOpen
  \bibfield  {author} {\bibinfo {author} {\bibfnamefont {E.}~\bibnamefont
  {Hudson}}, \bibinfo {author} {\bibfnamefont {D.~A.}\ \bibnamefont {Shirley}},
  \bibinfo {author} {\bibfnamefont {M.}~\bibnamefont {Domke}}, \bibinfo
  {author} {\bibfnamefont {G.}~\bibnamefont {Remmers}}, \bibinfo {author}
  {\bibfnamefont {A.}~\bibnamefont {Puschmann}}, \bibinfo {author}
  {\bibfnamefont {T.}~\bibnamefont {Mandel}}, \bibinfo {author} {\bibfnamefont
  {C.}~\bibnamefont {Xue}},\ and\ \bibinfo {author} {\bibfnamefont
  {G.}~\bibnamefont {Kaindl}},\ }\bibfield  {title} {\bibinfo {title}
  {High-resolution measurements of near-edge resonances in the core-level
  photoionization spectra of {SF}$_{6}$},\ }\href
  {https://doi.org/10.1103/PhysRevA.47.361} {\bibfield  {journal} {\bibinfo
  {journal} {Phys. Rev. A}\ }\textbf {\bibinfo {volume} {47}},\ \bibinfo
  {pages} {361} (\bibinfo {year} {1993})}\BibitemShut {NoStop}%
\bibitem [{\citenamefont {Stener}\ \emph {et~al.}(2011)\citenamefont {Stener},
  \citenamefont {Bolognesi}, \citenamefont {Coreno}, \citenamefont
  {O’Keeffe}, \citenamefont {Feyer}, \citenamefont {Fronzoni}, \citenamefont
  {Decleva}, \citenamefont {Avaldi},\ and\ \citenamefont
  {Kivim{\"a}ki}}]{stenerJCP2011}%
  \BibitemOpen
  \bibfield  {author} {\bibinfo {author} {\bibfnamefont {M.}~\bibnamefont
  {Stener}}, \bibinfo {author} {\bibfnamefont {P.}~\bibnamefont {Bolognesi}},
  \bibinfo {author} {\bibfnamefont {M.}~\bibnamefont {Coreno}}, \bibinfo
  {author} {\bibfnamefont {P.}~\bibnamefont {O’Keeffe}}, \bibinfo {author}
  {\bibfnamefont {V.}~\bibnamefont {Feyer}}, \bibinfo {author} {\bibfnamefont
  {G.}~\bibnamefont {Fronzoni}}, \bibinfo {author} {\bibfnamefont
  {P.}~\bibnamefont {Decleva}}, \bibinfo {author} {\bibfnamefont
  {L.}~\bibnamefont {Avaldi}},\ and\ \bibinfo {author} {\bibfnamefont
  {A.}~\bibnamefont {Kivim{\"a}ki}},\ }\bibfield  {title} {\bibinfo {title}
  {Photoabsorption and {S} 2p photoionization of the {SF$_6$} molecule:
  Resonances in the excitation energy range of 200--280 ev},\ }\href
  {https://doi.org/10.1063/1.3583815} {\bibfield  {journal} {\bibinfo
  {journal} {The Journal of chemical physics}\ }\textbf {\bibinfo {volume}
  {134}},\ \bibinfo {pages} {174311} (\bibinfo {year} {2011})}\BibitemShut
  {NoStop}%
\bibitem [{\citenamefont {Claassen}\ \emph {et~al.}(1970)\citenamefont
  {Claassen}, \citenamefont {Goodman}, \citenamefont {Holloway},\ and\
  \citenamefont {Selig}}]{claassenJCP1970}%
  \BibitemOpen
  \bibfield  {author} {\bibinfo {author} {\bibfnamefont {H.~H.}\ \bibnamefont
  {Claassen}}, \bibinfo {author} {\bibfnamefont {G.~L.}\ \bibnamefont
  {Goodman}}, \bibinfo {author} {\bibfnamefont {J.~H.}\ \bibnamefont
  {Holloway}},\ and\ \bibinfo {author} {\bibfnamefont {H.}~\bibnamefont
  {Selig}},\ }\bibfield  {title} {\bibinfo {title} {Raman spectra of {MoF$_6$},
  {TcF$_6$}, {ReF$_6$}, {UF$_6$}, {SF$_6$}, {SeF$_6$}, and {TeF$_6$} in the
  vapor state},\ }\href {https://doi.org/10.1063/1.1673786} {\bibfield
  {journal} {\bibinfo  {journal} {The Journal of Chemical Physics}\ }\textbf
  {\bibinfo {volume} {53}},\ \bibinfo {pages} {341} (\bibinfo {year}
  {1970})}\BibitemShut {NoStop}%
\bibitem [{\citenamefont {Wagner}\ \emph {et~al.}(2006)\citenamefont {Wagner},
  \citenamefont {W{\"u}est}, \citenamefont {Christov}, \citenamefont
  {Popmintchev}, \citenamefont {Zhou}, \citenamefont {Murnane},\ and\
  \citenamefont {Kapteyn}}]{wagnerPNAS2006}%
  \BibitemOpen
  \bibfield  {author} {\bibinfo {author} {\bibfnamefont {N.~L.}\ \bibnamefont
  {Wagner}}, \bibinfo {author} {\bibfnamefont {A.}~\bibnamefont {W{\"u}est}},
  \bibinfo {author} {\bibfnamefont {I.~P.}\ \bibnamefont {Christov}}, \bibinfo
  {author} {\bibfnamefont {T.}~\bibnamefont {Popmintchev}}, \bibinfo {author}
  {\bibfnamefont {X.}~\bibnamefont {Zhou}}, \bibinfo {author} {\bibfnamefont
  {M.~M.}\ \bibnamefont {Murnane}},\ and\ \bibinfo {author} {\bibfnamefont
  {H.~C.}\ \bibnamefont {Kapteyn}},\ }\bibfield  {title} {\bibinfo {title}
  {Monitoring molecular dynamics using coherent electrons from high harmonic
  generation},\ }\href {https://doi.org/10.1073/pnas.0605178103} {\bibfield
  {journal} {\bibinfo  {journal} {Proceedings of the National Academy of
  Sciences}\ }\textbf {\bibinfo {volume} {103}},\ \bibinfo {pages} {13279}
  (\bibinfo {year} {2006})}\BibitemShut {NoStop}%
\bibitem [{\citenamefont {Ferr{\'e}}\ \emph {et~al.}(2014)\citenamefont
  {Ferr{\'e}}, \citenamefont {Staedter}, \citenamefont {Burgy}, \citenamefont
  {Dagan}, \citenamefont {Descamps}, \citenamefont {Dudovich}, \citenamefont
  {Petit}, \citenamefont {Soifer}, \citenamefont {Blanchet},\ and\
  \citenamefont {Mairesse}}]{ferreJPB2014}%
  \BibitemOpen
  \bibfield  {author} {\bibinfo {author} {\bibfnamefont {A.}~\bibnamefont
  {Ferr{\'e}}}, \bibinfo {author} {\bibfnamefont {D.}~\bibnamefont {Staedter}},
  \bibinfo {author} {\bibfnamefont {F.}~\bibnamefont {Burgy}}, \bibinfo
  {author} {\bibfnamefont {M.}~\bibnamefont {Dagan}}, \bibinfo {author}
  {\bibfnamefont {D.}~\bibnamefont {Descamps}}, \bibinfo {author}
  {\bibfnamefont {N.}~\bibnamefont {Dudovich}}, \bibinfo {author}
  {\bibfnamefont {S.}~\bibnamefont {Petit}}, \bibinfo {author} {\bibfnamefont
  {H.}~\bibnamefont {Soifer}}, \bibinfo {author} {\bibfnamefont
  {V.}~\bibnamefont {Blanchet}},\ and\ \bibinfo {author} {\bibfnamefont
  {Y.}~\bibnamefont {Mairesse}},\ }\bibfield  {title} {\bibinfo {title}
  {High-order harmonic transient grating spectroscopy of {SF}$_6$ molecular
  vibrations},\ }\href {https://doi.org/10.1088/0953-4075/47/12/124023}
  {\bibfield  {journal} {\bibinfo  {journal} {Journal of Physics B: Atomic,
  Molecular and Optical Physics}\ }\textbf {\bibinfo {volume} {47}},\ \bibinfo
  {pages} {124023} (\bibinfo {year} {2014})}\BibitemShut {NoStop}%
\bibitem [{\citenamefont {Ferr{\'e}}\ \emph {et~al.}(2015)\citenamefont
  {Ferr{\'e}}, \citenamefont {Boguslavskiy}, \citenamefont {Dagan},
  \citenamefont {Blanchet}, \citenamefont {Bruner}, \citenamefont {Burgy},
  \citenamefont {Camper}, \citenamefont {Descamps}, \citenamefont {Fabre},
  \citenamefont {Fedorov} \emph {et~al.}}]{ferreNatComm2015}%
  \BibitemOpen
  \bibfield  {author} {\bibinfo {author} {\bibfnamefont {A.}~\bibnamefont
  {Ferr{\'e}}}, \bibinfo {author} {\bibfnamefont {A.}~\bibnamefont
  {Boguslavskiy}}, \bibinfo {author} {\bibfnamefont {M.}~\bibnamefont {Dagan}},
  \bibinfo {author} {\bibfnamefont {V.}~\bibnamefont {Blanchet}}, \bibinfo
  {author} {\bibfnamefont {B.}~\bibnamefont {Bruner}}, \bibinfo {author}
  {\bibfnamefont {F.}~\bibnamefont {Burgy}}, \bibinfo {author} {\bibfnamefont
  {A.}~\bibnamefont {Camper}}, \bibinfo {author} {\bibfnamefont
  {D.}~\bibnamefont {Descamps}}, \bibinfo {author} {\bibfnamefont
  {B.}~\bibnamefont {Fabre}}, \bibinfo {author} {\bibfnamefont
  {N.}~\bibnamefont {Fedorov}}, \emph {et~al.},\ }\bibfield  {title} {\bibinfo
  {title} {Multi-channel electronic and vibrational dynamics in polyatomic
  resonant high-order harmonic generation},\ }\href
  {https://doi.org/10.1038/ncomms6952} {\bibfield  {journal} {\bibinfo
  {journal} {Nature communications}\ }\textbf {\bibinfo {volume} {6}},\
  \bibinfo {pages} {1} (\bibinfo {year} {2015})}\BibitemShut {NoStop}%
\bibitem [{\citenamefont {Baykusheva}\ \emph {et~al.}(2016)\citenamefont
  {Baykusheva}, \citenamefont {Ahsan}, \citenamefont {Lin},\ and\ \citenamefont
  {W\"orner}}]{BaykushevaPRL2016}%
  \BibitemOpen
  \bibfield  {author} {\bibinfo {author} {\bibfnamefont {D.}~\bibnamefont
  {Baykusheva}}, \bibinfo {author} {\bibfnamefont {M.~S.}\ \bibnamefont
  {Ahsan}}, \bibinfo {author} {\bibfnamefont {N.}~\bibnamefont {Lin}},\ and\
  \bibinfo {author} {\bibfnamefont {H.~J.}\ \bibnamefont {W\"orner}},\
  }\bibfield  {title} {\bibinfo {title} {Bicircular high-harmonic spectroscopy
  reveals dynamical symmetries of atoms and molecules},\ }\href
  {https://doi.org/10.1103/PhysRevLett.116.123001} {\bibfield  {journal}
  {\bibinfo  {journal} {Phys. Rev. Lett.}\ }\textbf {\bibinfo {volume} {116}},\
  \bibinfo {pages} {123001} (\bibinfo {year} {2016})}\BibitemShut {NoStop}%
\bibitem [{\citenamefont {Hohenberg}\ and\ \citenamefont
  {Kohn}(1964)}]{HohenbergPR1964}%
  \BibitemOpen
  \bibfield  {author} {\bibinfo {author} {\bibfnamefont {P.}~\bibnamefont
  {Hohenberg}}\ and\ \bibinfo {author} {\bibfnamefont {W.}~\bibnamefont
  {Kohn}},\ }\bibfield  {title} {\bibinfo {title} {Inhomogeneous electron
  gas},\ }\href {https://doi.org/10.1103/PhysRev.136.B864} {\bibfield
  {journal} {\bibinfo  {journal} {Phys. Rev.}\ }\textbf {\bibinfo {volume}
  {136}},\ \bibinfo {pages} {B864} (\bibinfo {year} {1964})}\BibitemShut
  {NoStop}%
\bibitem [{\citenamefont {Runge}\ and\ \citenamefont
  {Gross}(1984)}]{RungePRL1984}%
  \BibitemOpen
  \bibfield  {author} {\bibinfo {author} {\bibfnamefont {E.}~\bibnamefont
  {Runge}}\ and\ \bibinfo {author} {\bibfnamefont {E.~K.~U.}\ \bibnamefont
  {Gross}},\ }\bibfield  {title} {\bibinfo {title} {Density-functional theory
  for time-dependent systems},\ }\href
  {https://doi.org/10.1103/PhysRevLett.52.997} {\bibfield  {journal} {\bibinfo
  {journal} {Phys. Rev. Lett.}\ }\textbf {\bibinfo {volume} {52}},\ \bibinfo
  {pages} {997} (\bibinfo {year} {1984})}\BibitemShut {NoStop}%
\bibitem [{\citenamefont {Neese}(2018)}]{NeeseORCA}%
  \BibitemOpen
  \bibfield  {author} {\bibinfo {author} {\bibfnamefont {F.}~\bibnamefont
  {Neese}},\ }\bibfield  {title} {\bibinfo {title} {Software update: the orca
  program system, version 4.0},\ }\href
  {https://doi.org/https://doi.org/10.1002/wcms.1327} {\bibfield  {journal}
  {\bibinfo  {journal} {WIREs Computational Molecular Science}\ }\textbf
  {\bibinfo {volume} {8}},\ \bibinfo {pages} {e1327} (\bibinfo {year}
  {2018})}\BibitemShut {NoStop}%
\bibitem [{\citenamefont {Becke}(1993)}]{BeckeJCP1993}%
  \BibitemOpen
  \bibfield  {author} {\bibinfo {author} {\bibfnamefont {A.~D.}\ \bibnamefont
  {Becke}},\ }\bibfield  {title} {\bibinfo {title} {Density‐functional
  thermochemistry. iii. the role of exact exchange},\ }\href
  {https://doi.org/10.1063/1.464913} {\bibfield  {journal} {\bibinfo  {journal}
  {The Journal of Chemical Physics}\ }\textbf {\bibinfo {volume} {98}},\
  \bibinfo {pages} {5648} (\bibinfo {year} {1993})}\BibitemShut {NoStop}%
\bibitem [{\citenamefont {Lee}\ \emph {et~al.}(1988)\citenamefont {Lee},
  \citenamefont {Yang},\ and\ \citenamefont {Parr}}]{LeePRB1988}%
  \BibitemOpen
  \bibfield  {author} {\bibinfo {author} {\bibfnamefont {C.}~\bibnamefont
  {Lee}}, \bibinfo {author} {\bibfnamefont {W.}~\bibnamefont {Yang}},\ and\
  \bibinfo {author} {\bibfnamefont {R.~G.}\ \bibnamefont {Parr}},\ }\bibfield
  {title} {\bibinfo {title} {Development of the colle-salvetti
  correlation-energy formula into a functional of the electron density},\
  }\href {https://doi.org/10.1103/PhysRevB.37.785} {\bibfield  {journal}
  {\bibinfo  {journal} {Phys. Rev. B}\ }\textbf {\bibinfo {volume} {37}},\
  \bibinfo {pages} {785} (\bibinfo {year} {1988})}\BibitemShut {NoStop}%
\bibitem [{\citenamefont {Vosko}\ \emph {et~al.}(1980)\citenamefont {Vosko},
  \citenamefont {Wilk},\ and\ \citenamefont {Nusair}}]{vosko1980}%
  \BibitemOpen
  \bibfield  {author} {\bibinfo {author} {\bibfnamefont {S.~H.}\ \bibnamefont
  {Vosko}}, \bibinfo {author} {\bibfnamefont {L.}~\bibnamefont {Wilk}},\ and\
  \bibinfo {author} {\bibfnamefont {M.}~\bibnamefont {Nusair}},\ }\bibfield
  {title} {\bibinfo {title} {Accurate spin-dependent electron liquid
  correlation energies for local spin density calculations: a critical
  analysis},\ }\href@noop {} {\bibfield  {journal} {\bibinfo  {journal}
  {Canadian Journal of physics}\ }\textbf {\bibinfo {volume} {58}},\ \bibinfo
  {pages} {1200} (\bibinfo {year} {1980})}\BibitemShut {NoStop}%
\bibitem [{\citenamefont {Neese}\ and\ \citenamefont
  {Valeev}(2011)}]{NeeseJCTC2011}%
  \BibitemOpen
  \bibfield  {author} {\bibinfo {author} {\bibfnamefont {F.}~\bibnamefont
  {Neese}}\ and\ \bibinfo {author} {\bibfnamefont {E.~F.}\ \bibnamefont
  {Valeev}},\ }\bibfield  {title} {\bibinfo {title} {Revisiting the atomic
  natural orbital approach for basis sets: Robust systematic basis sets for
  explicitly correlated and conventional correlated ab initio methods?},\
  }\href {https://doi.org/10.1021/ct100396y} {\bibfield  {journal} {\bibinfo
  {journal} {Journal of Chemical Theory and Computation}\ }\textbf {\bibinfo
  {volume} {7}},\ \bibinfo {pages} {33} (\bibinfo {year} {2011})},\ \bibinfo
  {note} {pMID: 26606216}\BibitemShut {NoStop}%
\bibitem [{\citenamefont {Kendall}\ and\ \citenamefont
  {Fr\"uchtl}(1997)}]{KendallFruchtl1997}%
  \BibitemOpen
  \bibfield  {author} {\bibinfo {author} {\bibfnamefont {R.}~\bibnamefont
  {Kendall}}\ and\ \bibinfo {author} {\bibfnamefont {H.}~\bibnamefont
  {Fr\"uchtl}},\ }\bibfield  {title} {\bibinfo {title} {The impact of the
  resolution of the identity approximate integral method on modern ab initio
  algorithm development},\ }\href {https://doi.org/10.1007/s002140050249}
  {\bibfield  {journal} {\bibinfo  {journal} {Theor Chem Acta}\ }\textbf
  {\bibinfo {volume} {97}},\ \bibinfo {pages} {158–163} (\bibinfo {year}
  {1997})}\BibitemShut {NoStop}%
\bibitem [{\citenamefont {de~Souza}\ \emph {et~al.}(2019)\citenamefont
  {de~Souza}, \citenamefont {Farias}, \citenamefont {Neese},\ and\
  \citenamefont {Izsák}}]{deSouzaJCTC2019}%
  \BibitemOpen
  \bibfield  {author} {\bibinfo {author} {\bibfnamefont {B.}~\bibnamefont
  {de~Souza}}, \bibinfo {author} {\bibfnamefont {G.}~\bibnamefont {Farias}},
  \bibinfo {author} {\bibfnamefont {F.}~\bibnamefont {Neese}},\ and\ \bibinfo
  {author} {\bibfnamefont {R.}~\bibnamefont {Izsák}},\ }\bibfield  {title}
  {\bibinfo {title} {Predicting phosphorescence rates of light organic
  molecules using time-dependent density functional theory and the path
  integral approach to dynamics},\ }\href
  {https://doi.org/10.1021/acs.jctc.8b00841} {\bibfield  {journal} {\bibinfo
  {journal} {Journal of Chemical Theory and Computation}\ }\textbf {\bibinfo
  {volume} {15}},\ \bibinfo {pages} {1896} (\bibinfo {year}
  {2019})}\BibitemShut {NoStop}%
\bibitem [{\citenamefont {Schmidt}\ \emph {et~al.}(1993)\citenamefont
  {Schmidt}, \citenamefont {Baldridge}, \citenamefont {Boatz}, \citenamefont
  {Elbert}, \citenamefont {Gordon}, \citenamefont {Jensen}, \citenamefont
  {Koseki}, \citenamefont {Matsunaga}, \citenamefont {Nguyen}, \citenamefont
  {Su}, \citenamefont {Windus}, \citenamefont {Dupuis},\ and\ \citenamefont
  {Montgomery~Jr}}]{Gamess}%
  \BibitemOpen
  \bibfield  {author} {\bibinfo {author} {\bibfnamefont {M.~W.}\ \bibnamefont
  {Schmidt}}, \bibinfo {author} {\bibfnamefont {K.~K.}\ \bibnamefont
  {Baldridge}}, \bibinfo {author} {\bibfnamefont {J.~A.}\ \bibnamefont
  {Boatz}}, \bibinfo {author} {\bibfnamefont {S.~T.}\ \bibnamefont {Elbert}},
  \bibinfo {author} {\bibfnamefont {M.~S.}\ \bibnamefont {Gordon}}, \bibinfo
  {author} {\bibfnamefont {J.~H.}\ \bibnamefont {Jensen}}, \bibinfo {author}
  {\bibfnamefont {S.}~\bibnamefont {Koseki}}, \bibinfo {author} {\bibfnamefont
  {N.}~\bibnamefont {Matsunaga}}, \bibinfo {author} {\bibfnamefont {K.~A.}\
  \bibnamefont {Nguyen}}, \bibinfo {author} {\bibfnamefont {S.}~\bibnamefont
  {Su}}, \bibinfo {author} {\bibfnamefont {T.~L.}\ \bibnamefont {Windus}},
  \bibinfo {author} {\bibfnamefont {M.}~\bibnamefont {Dupuis}},\ and\ \bibinfo
  {author} {\bibfnamefont {J.~A.}\ \bibnamefont {Montgomery~Jr}},\ }\bibfield
  {title} {\bibinfo {title} {General atomic and molecular electronic structure
  system},\ }\href {https://doi.org/https://doi.org/10.1002/jcc.540141112}
  {\bibfield  {journal} {\bibinfo  {journal} {Journal of Computational
  Chemistry}\ }\textbf {\bibinfo {volume} {14}},\ \bibinfo {pages} {1347}
  (\bibinfo {year} {1993})}\BibitemShut {NoStop}%
\bibitem [{\citenamefont {Weigend}\ and\ \citenamefont
  {Ahlrichs}(2005)}]{WeigendPCCP2005}%
  \BibitemOpen
  \bibfield  {author} {\bibinfo {author} {\bibfnamefont {F.}~\bibnamefont
  {Weigend}}\ and\ \bibinfo {author} {\bibfnamefont {R.}~\bibnamefont
  {Ahlrichs}},\ }\bibfield  {title} {\bibinfo {title} {Balanced basis sets of
  split valence{,} triple zeta valence and quadruple zeta valence quality for h
  to rn: Design and assessment of accuracy},\ }\href
  {https://doi.org/10.1039/B508541A} {\bibfield  {journal} {\bibinfo  {journal}
  {Phys. Chem. Chem. Phys.}\ }\textbf {\bibinfo {volume} {7}},\ \bibinfo
  {pages} {3297} (\bibinfo {year} {2005})}\BibitemShut {NoStop}%
\bibitem [{\citenamefont {Barreau}\ \emph {et~al.}(2020)\citenamefont
  {Barreau}, \citenamefont {Ross}, \citenamefont {Garg}, \citenamefont {Kraus},
  \citenamefont {Neumark},\ and\ \citenamefont {Leone}}]{BarreauSciRep2020}%
  \BibitemOpen
  \bibfield  {author} {\bibinfo {author} {\bibfnamefont {L.}~\bibnamefont
  {Barreau}}, \bibinfo {author} {\bibfnamefont {A.~D.}\ \bibnamefont {Ross}},
  \bibinfo {author} {\bibfnamefont {S.}~\bibnamefont {Garg}}, \bibinfo {author}
  {\bibfnamefont {P.~M.}\ \bibnamefont {Kraus}}, \bibinfo {author}
  {\bibfnamefont {D.~M.}\ \bibnamefont {Neumark}},\ and\ \bibinfo {author}
  {\bibfnamefont {S.~R.}\ \bibnamefont {Leone}},\ }\bibfield  {title} {\bibinfo
  {title} {Efficient table-top dual-wavelength beamline for ultrafast transient
  absorption spectroscopy in the soft {X}-ray region},\ }\href
  {https://doi.org/https://doi.org/10.1038/s41598-020-62461-6} {\bibfield
  {journal} {\bibinfo  {journal} {Scientific reports}\ }\textbf {\bibinfo
  {volume} {10}},\ \bibinfo {pages} {1} (\bibinfo {year} {2020})}\BibitemShut
  {NoStop}%
\bibitem [{\citenamefont {G{\'e}neaux}\ \emph {et~al.}(2021)\citenamefont
  {G{\'e}neaux}, \citenamefont {Chang}, \citenamefont {Schwartzberg},\ and\
  \citenamefont {Marroux}}]{GeneauxOE2021}%
  \BibitemOpen
  \bibfield  {author} {\bibinfo {author} {\bibfnamefont {R.}~\bibnamefont
  {G{\'e}neaux}}, \bibinfo {author} {\bibfnamefont {H.-T.}\ \bibnamefont
  {Chang}}, \bibinfo {author} {\bibfnamefont {A.~M.}\ \bibnamefont
  {Schwartzberg}},\ and\ \bibinfo {author} {\bibfnamefont {H.~J.}\ \bibnamefont
  {Marroux}},\ }\bibfield  {title} {\bibinfo {title} {Source noise suppression
  in attosecond transient absorption spectroscopy by edge-pixel referencing},\
  }\href {https://doi.org/https://doi.org/10.1364/OE.412117} {\bibfield
  {journal} {\bibinfo  {journal} {Optics Express}\ }\textbf {\bibinfo {volume}
  {29}},\ \bibinfo {pages} {951} (\bibinfo {year} {2021})}\BibitemShut
  {NoStop}%
\bibitem [{\citenamefont {Nguyen}\ \emph {et~al.}(2016)\citenamefont {Nguyen},
  \citenamefont {Lucchese}, \citenamefont {Lin},\ and\ \citenamefont
  {Le}}]{NguyenPRA2016}%
  \BibitemOpen
  \bibfield  {author} {\bibinfo {author} {\bibfnamefont {N.-T.}\ \bibnamefont
  {Nguyen}}, \bibinfo {author} {\bibfnamefont {R.~R.}\ \bibnamefont
  {Lucchese}}, \bibinfo {author} {\bibfnamefont {C.~D.}\ \bibnamefont {Lin}},\
  and\ \bibinfo {author} {\bibfnamefont {A.-T.}\ \bibnamefont {Le}},\
  }\bibfield  {title} {\bibinfo {title} {Probing and extracting the structure
  of vibrating {SF}$_{6}$ molecules with inner-shell photoelectrons},\ }\href
  {https://doi.org/10.1103/PhysRevA.93.063419} {\bibfield  {journal} {\bibinfo
  {journal} {Phys. Rev. A}\ }\textbf {\bibinfo {volume} {93}},\ \bibinfo
  {pages} {063419} (\bibinfo {year} {2016})}\BibitemShut {NoStop}%
\bibitem [{\citenamefont {Rupprecht}\ \emph {et~al.}(2022)\citenamefont
  {Rupprecht}, \citenamefont {Aufleger}, \citenamefont {Heinze}, \citenamefont
  {Magunia}, \citenamefont {Ding}, \citenamefont {Rebholz}, \citenamefont
  {Amberg}, \citenamefont {Mollov}, \citenamefont {Henrich}, \citenamefont
  {Haverkort}, \citenamefont {Ott},\ and\ \citenamefont
  {Pfeifer}}]{RupprechtPRL2022}%
  \BibitemOpen
  \bibfield  {author} {\bibinfo {author} {\bibfnamefont {P.}~\bibnamefont
  {Rupprecht}}, \bibinfo {author} {\bibfnamefont {L.}~\bibnamefont {Aufleger}},
  \bibinfo {author} {\bibfnamefont {S.}~\bibnamefont {Heinze}}, \bibinfo
  {author} {\bibfnamefont {A.}~\bibnamefont {Magunia}}, \bibinfo {author}
  {\bibfnamefont {T.}~\bibnamefont {Ding}}, \bibinfo {author} {\bibfnamefont
  {M.}~\bibnamefont {Rebholz}}, \bibinfo {author} {\bibfnamefont
  {S.}~\bibnamefont {Amberg}}, \bibinfo {author} {\bibfnamefont
  {N.}~\bibnamefont {Mollov}}, \bibinfo {author} {\bibfnamefont
  {F.}~\bibnamefont {Henrich}}, \bibinfo {author} {\bibfnamefont {M.~W.}\
  \bibnamefont {Haverkort}}, \bibinfo {author} {\bibfnamefont {C.}~\bibnamefont
  {Ott}},\ and\ \bibinfo {author} {\bibfnamefont {T.}~\bibnamefont {Pfeifer}},\
  }\bibfield  {title} {\bibinfo {title} {Laser control of electronic exchange
  interaction within a molecule},\ }\href
  {https://doi.org/10.1103/PhysRevLett.128.153001} {\bibfield  {journal}
  {\bibinfo  {journal} {Phys. Rev. Lett.}\ }\textbf {\bibinfo {volume} {128}},\
  \bibinfo {pages} {153001} (\bibinfo {year} {2022})}\BibitemShut {NoStop}%
\bibitem [{\citenamefont {Pertot}\ \emph {et~al.}(2017)\citenamefont {Pertot},
  \citenamefont {Schmidt}, \citenamefont {Matthews}, \citenamefont {Chauvet},
  \citenamefont {Huppert}, \citenamefont {Svoboda}, \citenamefont {von Conta},
  \citenamefont {Tehlar}, \citenamefont {Baykusheva}, \citenamefont {Wolf}
  \emph {et~al.}}]{pertotScience2017}%
  \BibitemOpen
  \bibfield  {author} {\bibinfo {author} {\bibfnamefont {Y.}~\bibnamefont
  {Pertot}}, \bibinfo {author} {\bibfnamefont {C.}~\bibnamefont {Schmidt}},
  \bibinfo {author} {\bibfnamefont {M.}~\bibnamefont {Matthews}}, \bibinfo
  {author} {\bibfnamefont {A.}~\bibnamefont {Chauvet}}, \bibinfo {author}
  {\bibfnamefont {M.}~\bibnamefont {Huppert}}, \bibinfo {author} {\bibfnamefont
  {V.}~\bibnamefont {Svoboda}}, \bibinfo {author} {\bibfnamefont
  {A.}~\bibnamefont {von Conta}}, \bibinfo {author} {\bibfnamefont
  {A.}~\bibnamefont {Tehlar}}, \bibinfo {author} {\bibfnamefont
  {D.}~\bibnamefont {Baykusheva}}, \bibinfo {author} {\bibfnamefont {J.-P.}\
  \bibnamefont {Wolf}}, \emph {et~al.},\ }\bibfield  {title} {\bibinfo {title}
  {Time-resolved {X}-ray absorption spectroscopy with a water window
  high-harmonic source},\ }\href {https://doi.org/10.1126/science.aah6114}
  {\bibfield  {journal} {\bibinfo  {journal} {Science}\ }\textbf {\bibinfo
  {volume} {355}},\ \bibinfo {pages} {264} (\bibinfo {year}
  {2017})}\BibitemShut {NoStop}%
\bibitem [{\citenamefont {Ross}\ \emph {et~al.}(2022)\citenamefont {Ross},
  \citenamefont {Hait}, \citenamefont {Scutelnic}, \citenamefont {Haugen},
  \citenamefont {Ridente}, \citenamefont {Balkew}, \citenamefont {Neumark},
  \citenamefont {Head-Gordon},\ and\ \citenamefont {Leone}}]{RossArXiv2022}%
  \BibitemOpen
  \bibfield  {author} {\bibinfo {author} {\bibfnamefont {A.~D.}\ \bibnamefont
  {Ross}}, \bibinfo {author} {\bibfnamefont {D.}~\bibnamefont {Hait}}, \bibinfo
  {author} {\bibfnamefont {V.}~\bibnamefont {Scutelnic}}, \bibinfo {author}
  {\bibfnamefont {E.~A.}\ \bibnamefont {Haugen}}, \bibinfo {author}
  {\bibfnamefont {E.}~\bibnamefont {Ridente}}, \bibinfo {author} {\bibfnamefont
  {M.~B.}\ \bibnamefont {Balkew}}, \bibinfo {author} {\bibfnamefont {D.~M.}\
  \bibnamefont {Neumark}}, \bibinfo {author} {\bibfnamefont {M.}~\bibnamefont
  {Head-Gordon}},\ and\ \bibinfo {author} {\bibfnamefont {S.~R.}\ \bibnamefont
  {Leone}},\ }\bibfield  {title} {\bibinfo {title} {Jahn-{T}eller distortion
  and dissociation of {CCl$_4^+$} by transient {X}-ray spectroscopy
  simultaneously at the carbon {K}- and chlorine {L}-edge},\ }\bibfield
  {journal} {\bibinfo  {journal} {arXiv}\ }\href
  {https://doi.org/10.48550/ARXIV.2204.13800} {10.48550/ARXIV.2204.13800}
  (\bibinfo {year} {2022})\BibitemShut {NoStop}%
\bibitem [{See()}]{SeeSM}%
  \BibitemOpen
  \href@noop {} {}\bibinfo {note} {See Supplemental Material}\BibitemShut
  {NoStop}%
\bibitem [{\citenamefont {Kivim{\"a}ki}\ \emph {et~al.}(2016)\citenamefont
  {Kivim{\"a}ki}, \citenamefont {Coreno}, \citenamefont {Miotti}, \citenamefont
  {Frassetto}, \citenamefont {Poletto}, \citenamefont {Str{\aa}hlman},
  \citenamefont {De~Simone},\ and\ \citenamefont
  {Richter}}]{kivimakiJESRP2016}%
  \BibitemOpen
  \bibfield  {author} {\bibinfo {author} {\bibfnamefont {A.}~\bibnamefont
  {Kivim{\"a}ki}}, \bibinfo {author} {\bibfnamefont {M.}~\bibnamefont
  {Coreno}}, \bibinfo {author} {\bibfnamefont {P.}~\bibnamefont {Miotti}},
  \bibinfo {author} {\bibfnamefont {F.}~\bibnamefont {Frassetto}}, \bibinfo
  {author} {\bibfnamefont {L.}~\bibnamefont {Poletto}}, \bibinfo {author}
  {\bibfnamefont {C.}~\bibnamefont {Str{\aa}hlman}}, \bibinfo {author}
  {\bibfnamefont {M.}~\bibnamefont {De~Simone}},\ and\ \bibinfo {author}
  {\bibfnamefont {R.}~\bibnamefont {Richter}},\ }\bibfield  {title} {\bibinfo
  {title} {The multielectron character of the {S} 2p$\rightarrow$ 4eg shape
  resonance in the sf$_6$ molecule studied via detection of soft{X}-ray
  emission and neutral high-rydberg fragments},\ }\href
  {https://doi.org/10.1016/j.elspec.2016.03.002} {\bibfield  {journal}
  {\bibinfo  {journal} {Journal of Electron Spectroscopy and Related
  Phenomena}\ }\textbf {\bibinfo {volume} {209}},\ \bibinfo {pages} {26}
  (\bibinfo {year} {2016})}\BibitemShut {NoStop}%
\end{thebibliography}%

\end{document}